\documentclass[a4paper,fleqn]{cas-sc}

\usepackage[authoryear,longnamesfirst]{natbib}

\def\tsc#1{\csdef{#1}{\textsc{\lowercase{#1}}\xspace}}
\tsc{WGM}
\tsc{QE}

\usepackage{graphicx}
\usepackage{subcaption}
\usepackage{multirow}
\usepackage{placeins}
\usepackage{algorithm}
\usepackage{algpseudocode}

\begin{document}
\let\WriteBookmarks\relax
\def\floatpagepagefraction{1}
\def\textpagefraction{.001}

\shorttitle{}    

\shortauthors{}  

\title [mode = title]{A coupled Eulerian–Lagrangian approach for fluid and particle dynamics}  

\tnotemark[1] 

\tnotetext[1]{} 

%

\author[1]{Snehanshu Maiti}

\cormark[1]


\ead{snehanshu.maiti@gmail.com}


\credit{}

\affiliation[1]{organization={Institute for Plasma Research},
            addressline={Bhat}, 
            city={Gandhinagar},
            postcode={382428}, 
            state={Gujarat},
            country={India}}

\author[1,2]{Rajaraman Ganesh}

\affiliation[2]{
organization={Homi Bhabha National Institute},
addressline={Training School Complex, Anushaktinagar},
city={Mumbai},
postcode={400094},
country={India}
}

\cortext[1]{Corresponding author}



\begin{abstract}
We present a modular one-way coupled Eulerian--Lagrangian computational framework for the simultaneous simulation of fluid and particle dynamics in two-dimensional incompressible flows. The framework extends the GPU-accelerated GHD2D Fourier pseudospectral Navier--Stokes solver \textcolor{black}{[\cite{Mukherjee2018,Biswas2024}]} by incorporating passive tracer and finite-inertia particle modules within a unified computational environment, enabling the evolution of large particle ensembles in dynamically evolving turbulent flows. The Eulerian fluid equations are integrated using a second-order Adams--Bashforth scheme, while particle trajectories are advanced with a classical fourth-order Runge--Kutta method. Coupling between the Eulerian and Lagrangian descriptions is achieved through spatial and temporal interpolation of the fluid quantities, with support for bilinear, bicubic Catmull--Rom, and bicubic B-spline interpolation schemes.
The numerical framework is systematically verified through analytical solutions and benchmark problems for the Eulerian fluid solver, tracer-particle transport, and inertial-particle dynamics. Interpolation studies demonstrate that bilinear interpolation provides transport statistics nearly identical to those obtained with higher-order schemes while offering superior computational efficiency. Particle-number convergence studies further establish the statistical robustness of the computed Lagrangian transport measures. The capabilities of the framework are demonstrated through simulations of tracer and inertial particles in decaying two-dimensional incompressible turbulence. The coupled solver accurately captures long-time particle transport, turbulent dispersion, vortex trapping, coherent-structure interactions, preferential concentration, and inertia-dependent transport over a broad range of flow conditions. Performance measurements show stable scaling with both Eulerian grid resolution and particle number, while maintaining efficient production-scale simulations on a single GPU. Owing to its modular software architecture, numerical robustness, and computational efficiency, the developed framework provides a flexible and extensible platform for future Eulerian--Lagrangian simulations of turbulent transport and particle-laden incompressible flows.

\end{abstract}

\maketitle

\section{Introduction}\label{}

The transport of particles by fluids is a fundamental problem encountered in a broad range of scientific applications, including atmospheric [\cite{Hadjighasem2014}] and oceanic transport [\cite{Corrado2017}],  cloud microphysics [\cite{Li2017,Ravichandran2015}], and astrophysical problems such as planetesimal formation in protoplanetary disks [\cite{Gerosa2023}]. While the evolution of the \textcolor{black}{fluid} is naturally described within an Eulerian framework, many transport phenomena are more appropriately characterized from a Lagrangian perspective by following the trajectories of individual particles embedded within the flow. Consequently, coupled Eulerian--Lagrangian approaches have become important \textcolor{black}{as} a computational tool for investigating particle transport, turbulent dispersion, mixing, coherent-structure interactions, and inertia-driven clustering in complex fluid flows.

The Eulerian description provides detailed information on the spatial and temporal evolution of flow quantities such as velocity, pressure, and vorticity. However, many physically important transport processes depend not only on the instantaneous flow field but also on the  trajectories of fluid \textcolor{black}{elements} and particles. Lagrangian particle tracking therefore complements the Eulerian description by providing direct access to quantities such as particle dispersion, residence times, preferential concentration, and long-time transport statistics. Depending on the physical problem, particles may be modeled either as \textcolor{black}{inertia-less} passive tracers that follow the local fluid velocity exactly or as finite-inertia particles whose motion deviates from fluid streamlines because of their finite response time.

A key challenge in coupled Eulerian--Lagrangian simulations is the accurate transfer of information between the Eulerian mesh and the Lagrangian particles. Since particle positions generally do not coincide with the fixed computational grid, the carrier-flow quantities required by the particle solver must be obtained through interpolation. The choice of interpolation strategy can influence particle trajectories and transport statistics, particularly in long-duration simulations where interpolation errors may accumulate over many integration steps. Consequently, systematic verification, interpolation assessment, and convergence studies are essential for establishing the numerical accuracy and robustness of Eulerian--Lagrangian frameworks intended for quantitative transport investigations.

The present work builds upon the GPU-accelerated GHD2D pseudospectral solver for two-dimensional incompressible Navier--Stokes turbulence, which has previously been employed to investigate hydrodynamic instabilities, vortex dynamics, coherent-structure formation, and turbulent evolution \textcolor{black}{[\cite{Mukherjee2018,Rupakthesis,BG2022,Biswas2024}].} While the original implementation provides an efficient and accurate Eulerian description of incompressible flows, it does not include a dedicated Lagrangian particle-tracking capability. Consequently, studies of particle transport, turbulent dispersion, preferential concentration, and inertial-particle dynamics could not be performed within the original computational framework.

In this work, we develop a modular one-way coupled Eulerian--Lagrangian framework, in which the fluid influences the particle motion while particle feedback on the fluid is neglected, by extending the GHD2D solver with passive tracer and finite-inertia particle modules.
The framework combines a Fourier pseudospectral Eulerian fluid solver with fourth-order Runge--Kutta particle integration through a unified spatial and temporal interpolation infrastructure, enabling the simultaneous evolution of dynamically evolving flow fields and large particle ensembles. In addition to the numerical algorithms, the implementation incorporates a modular software architecture and heterogeneous GPU--CPU execution strategy that facilitate future extensions while preserving compatibility with the underlying Eulerian solver. The framework supports both passive tracers and heavy inertial particles \textcolor{black}{governed by the simplified Maxey--Riley equation~[\cite{Maxey1983}]} under the Stokes-drag approximation, thereby enabling investigations of both fluid-element transport and finite-inertia particle dynamics within a common computational environment.

The numerical framework is systematically verified through analytical solutions and benchmark problems for the Eulerian fluid solver, tracer-particle integration, and inertial-particle dynamics. The influence of different interpolation strategies is assessed through dedicated interpolation studies, while particle-number convergence tests establish the statistical robustness of the computed transport measures. The capabilities of the framework are subsequently demonstrated through applications to decaying two-dimensional incompressible turbulence, including long-time tracer transport, turbulent dispersion, vortex trapping, preferential concentration, and inertia-dependent particle dynamics. Performance analyses further quantify the computational scaling of the coupled implementation with both Eulerian grid resolution and particle number. Beyond the applications considered here, the developed framework is intended to provide a flexible and extensible computational foundation for future Eulerian--Lagrangian investigations involving more sophisticated particle models, coupling strategies, and fluid solvers.

The remainder of this paper is organized as follows. Section~2 presents the governing equations for the Eulerian fluid solver and the Lagrangian particle models. Section~3 describes the numerical methodology, including the fluid solver, particle integrators, interpolation procedures, and coupled Eulerian--Lagrangian algorithm. Section~4 outlines the software architecture and parallel implementation. Section~5 presents the verification, benchmark and convergence studies. Applications to tracer and inertial particle transport in decaying two-dimensional turbulence are presented in Section~6. Section~7 discusses the computational performance of the coupled framework, and the conclusions are given in Section~8.

\section{Governing Equations}\label{}

The present framework combines an Eulerian description of the fluid dynamics with a Lagrangian description of passive tracer and inertial particles. The governing equations describing the fluid and particle dynamics are presented below.

\subsection{Fluid dynamics}

The fluid dynamics of decaying two-dimensional incompressible Navier--Stokes turbulence is described using the vorticity--streamfunction $(\omega,\psi)$ formulation,

\begin{equation}
\frac{\partial \omega}{\partial t}
=
[\psi,\omega]
+
\frac{1}{\mathrm{Re}}\nabla^2\omega,
\qquad
\omega = -\nabla^2\psi,
\qquad
[\psi,\omega]
=
\frac{\partial \psi}{\partial x}
\frac{\partial \omega}{\partial y}
-
\frac{\partial \psi}{\partial y}
\frac{\partial \omega}{\partial x}.
\label{eq:navier_stokes_scalar}
\end{equation}

Here, $[\psi,\omega]$ denotes the Poisson bracket, which represents the nonlinear advection of vorticity by the incompressible flow, and $\mathrm{Re}$ is the Reynolds number, representing the ratio of inertial to viscous forces. Larger values of {Re} correspond to increasingly turbulent flow conditions and weaker viscous dissipation. All quantities are expressed in nondimensional form.

\subsection{Tracer Particles}
Passive tracer particles are represented in a Lagrangian framework and are assumed to move exactly with the local fluid velocity. Their motion is governed by,

\begin{equation}
\frac{d\mathbf{x}}{dt}
=
\mathbf{u}(\mathbf{x},t),
\qquad
\mathbf{u}
=
\left(
\frac{\partial\psi}{\partial y},
-\frac{\partial\psi}{\partial x}
\right)
\label{eq:adv_eqn}
\end{equation}

Here, \(\mathbf{x}(t)=(x(t),y(t))\) denotes the instantaneous tracer-particle position and \(\mathbf{u}(\mathbf{x},t)\) is the fluid velocity evaluated at the particle location. The velocity field is obtained from the stream function \(\psi(x,y,t)\) through the incompressibility condition.

\subsection{Inertial Particles}

The dynamics of small, heavy inertial particles, which possess finite mass and respond to the fluid with a characteristic relaxation time, \textcolor{black}{are described by the simplified Maxey--Riley equation [\cite{Maxey1983}],}

\begin{equation}
\frac{d\mathbf{x}_p}{dt} = \mathbf{v}_p, \quad 
\frac{d\mathbf{v}_p}{dt} = \frac{1}{\mathrm{St}} \Big[\mathbf{u}(\mathbf{x}_p,t) - \mathbf{v}_p \Big],
\label{eq:inertial_particle}
\end{equation}

\hspace{-0.57cm} where \( \mathbf{x}_p(t) = (x_p(t), y_p(t)) \) and \( \mathbf{v}_p(t) = (v_{px}(t), v_{py}(t)) \) are the instantaneous position and velocity of the particle, respectively, \( \mathbf{u}(\mathbf{x}_p,t) = (u(x_p,y_p,t), v(x_p,y_p,t)) \) is the fluid velocity interpolated to the particle location, \textcolor{black}{and \( \mathrm{St} = \tau_p/\tau_f \) is the Stokes number, defined as the ratio of the particle response time ($\tau_p$) to the characteristic fluid timescale ($\tau_f$).} Thus, \( \mathrm{St} \) quantifies the particle inertia and determines how rapidly the particle velocity relaxes toward the local fluid velocity. \\

The governing equations presented above define the mathematical formulation of the coupled Eulerian--Lagrangian framework. The numerical methods employed for their spatial discretization, temporal integration, and coupling are described in the following section.

\section{Numerical Method}\label{}

\textcolor{black}{This section describes the numerical framework employed in the present Eulerian--Lagrangian solver. We first present the GPU-accelerated Eulerian fluid solver for the incompressible Navier--Stokes equations, followed by the numerical integration procedures for tracer and inertial particles. The implemented spatial and temporal interpolation schemes are then described, and the section concludes with the coupled Eulerian--Lagrangian integration algorithm that combines the fluid and particle solvers within a one-way coupling framework.}

\subsection{Eulerian fluid solver}

The Eulerian fluid dynamics are computed using the GPU-accelerated two-dimensional incompressible hydrodynamic solver GHD2D \textcolor{black}{[\cite{Rupakthesis,BG2022,Biswas2024}]}, developed in-house at the Institute for Plasma Research. The \textcolor{black}{GPU based} solver advances the vorticity--streamfunction formulation of the incompressible Navier--Stokes equations introduced in Sec.~2 on a doubly periodic square domain in Cartesian coordinates. GHD2D has been employed for high-resolution simulations, including grids of up to $2048^2$, and for flows at very high Reynolds numbers.

The spatial discretization is based on a Fourier pseudospectral method.  The vorticity field is represented through its Fourier coefficients $\hat{\omega}(\mathbf{k},t)$, where
$\mathbf{k}=(k_x,k_y)$ denotes the Fourier wave vector.
Linear operations, including spatial differentiation, the spectral
solution of the Poisson equation, and viscous dissipation, are carried
out directly in Fourier space, where they reduce to simple algebraic
multiplications or divisions. In contrast, the nonlinear advection
terms are evaluated in physical space to avoid the expensive
convolution sums that would arise from direct multiplication of Fourier series. Consequently, forward and inverse fast Fourier transforms are performed at each timestep to transfer the solution efficiently between the spectral and physical representations. Spatial derivatives are evaluated spectrally according to

\begin{equation}
\widehat{\frac{\partial \omega}{\partial x}}
=
\mathrm{i}k_x\hat{\omega},
\qquad
\widehat{\frac{\partial \omega}{\partial y}}
=
\mathrm{i}k_y\hat{\omega}.
\label{eq:spectral_vorticity_derivatives}
\end{equation}

\hspace{-0.52cm}\textcolor{black}{Using} the convention
$\omega=-\nabla^2\psi$, the stream function is obtained by solving the
Poisson equation independently for each nonzero Fourier mode,

\begin{equation}
\hat{\psi}(\mathbf{k},t)
=
\frac{\hat{\omega}(\mathbf{k},t)}
{k_x^2+k_y^2},
\qquad
\mathbf{k}\neq\mathbf{0},
\label{eq:poisson_spectral}
\end{equation}

\hspace{-0.52cm}\textcolor{black}{while} the zero mode is set to $\hat{\psi}(\mathbf{0},t)=0$. The velocity components are subsequently reconstructed in spectral space from the stream-function relations,

\begin{equation}
\hat{u}(\mathbf{k},t)
=
\mathrm{i}k_y\hat{\psi}(\mathbf{k},t),
\qquad
\hat{v}(\mathbf{k},t)
=
-\mathrm{i}k_x\hat{\psi}(\mathbf{k},t).
\label{eq:spectral_velocity}
\end{equation}

\hspace{-0.52cm}\textcolor{black}{The nonlinear} vorticity-advection term is evaluated using the standard
pseudospectral procedure. First, the spectral velocity components and
spectral vorticity derivatives are transformed to physical space using
inverse fast Fourier transforms. The nonlinear products

\begin{equation}
u\frac{\partial\omega}{\partial x},
\qquad
v\frac{\partial\omega}{\partial y}
\end{equation}

\hspace{-0.52cm}\textcolor{black}{are} then formed pointwise on the Eulerian grid. These products are
subsequently transformed back to Fourier space using forward FFTs, so
that the spectral nonlinear term is written as

\begin{equation}
\widehat{\mathcal{N}}(\mathbf{k},t)
=
\widehat{
u\frac{\partial\omega}{\partial x}
}
+
\widehat{
v\frac{\partial\omega}{\partial y}
}.
\label{eq:spectral_nonlinear_term}
\end{equation}

\hspace{-0.52cm}\textcolor{black}{To} suppress aliasing errors generated by the physical-space
multiplication, the standard $2/3$ dealiasing rule
\textcolor{black}{[\cite{Patterson1971}]} is applied to the transformed nonlinear terms,
with the highest one-third of the resolved spectral modes set to zero.
The viscous term is evaluated directly in Fourier space. The resulting
spectral evolution equation is therefore

\begin{equation}
\frac{\partial\hat{\omega}}{\partial t}
=
-\widehat{\mathcal{N}}(\mathbf{k},t)
-
\left(
\nu_x k_x^2+\nu_y k_y^2
\right)
\hat{\omega}(\mathbf{k},t),
\label{eq:spectral_vorticity_evolution}
\end{equation}

\hspace{-0.52cm}\textcolor{black}{where} $\nu_x$ and $\nu_y$ are the viscosity coefficients in the two
Cartesian directions. For the isotropic simulations considered here,
$\nu_x=\nu_y=\nu$.

Temporal integration is carried out entirely in spectral space using a
second-order Adams--Bashforth scheme
\textcolor{black}{[\cite{Adams1855,Bashforth1883}]}. Denoting the right-hand side of
Eq.~\eqref{eq:spectral_vorticity_evolution} by
$\mathcal{R}^n(\mathbf{k})$, the update is

\begin{equation}
\hat{\omega}^{\,n+1}
=
\hat{\omega}^{\,n}
+
\Delta t
\left[
\frac{3}{2}\mathcal{R}^{n}
-
\frac{1}{2}\mathcal{R}^{n-1}
\right],
\label{eq:ab2_fluid}
\end{equation}

where $\Delta t$ is the fluid-integration timestep. After each AB2
update, the new stream-function and velocity fields are reconstructed
from $\hat{\omega}^{\,n+1}$ using
Eqs.~\eqref{eq:poisson_spectral} and
\eqref{eq:spectral_velocity}. Inverse FFTs are then applied to recover
$\omega^{n+1}$, $\psi^{n+1}$, $u^{n+1}$, and $v^{n+1}$ on the physical
grid. The inverse-transformed fields are normalized by the total number
of grid points, $N_xN_y$.

All Fourier transforms are performed using the NVIDIA cuFFT library
\textcolor{black}{[\cite{nvidia_cufft}]}. Double-precision real-to-complex transforms are
used for forward transformations of physical-space fields, whereas
double-complex-to-real transforms are used to recover the corresponding
Eulerian fields. OpenACC directives are employed to execute the
spectral operations, pointwise nonlinear products, normalization,
diagnostic reductions, and field updates on a single NVIDIA GPU. The
cuFFT plans are created once at the beginning of the simulation and
reused throughout the entire time integration, thereby avoiding
repeated plan-construction overhead.

The fluid solver constitutes the Eulerian component of the coupled Eulerian–Lagrangian framework. At each global timestep, the vorticity field is first advanced from $t_n$ to $t_{n+1}$, after which the updated stream-function and velocity fields are reconstructed. The Eulerian fields at the two successive time levels are then supplied to the Lagrangian particle module for spatial and temporal interpolation during the RK4 particle update. Unless otherwise stated, the coupled simulations use a common timestep of
$\Delta t=\Delta t_p=10^{-3}$ for both the Eulerian and Lagrangian
integrators.

\subsection{Tracer-particle integration}


The tracer-particle equations introduced in Sec.~2 are integrated using a classical fourth-order Runge--Kutta (RK4) scheme, which provides fourth-order global accuracy and robust long-time trajectory integration.

Since particle positions generally do not coincide with Eulerian grid locations, the fluid quantities required by the particle solver are obtained through interpolation. The present framework supports particle integration using either direct interpolation of the velocity field or interpolation of the stream-function field. Throughout the turbulent simulations reported in this work, the latter approach is adopted.

At each Runge--Kutta stage, the intermediate particle position is first predicted using the velocity obtained from the preceding stage.
The four RK4 stages are evaluated at the times $t_n$, $t_n+\Delta t_p/2$, $t_n+\Delta t_p/2$, and $t_n+\Delta t_p$, where $\Delta t_p$ denotes the particle-integration time step. Before each interpolation operation, the four neighboring points required for the centered finite-difference evaluation, $(x_p\pm\Delta x,y_p)$ and $(x_p,y_p\pm\Delta y)$, are constructed around the intermediate RK4 particle position. If any of these points lies outside the computational domain, it is mapped periodically to the opposite side of the domain before the stream-function interpolation is performed.
The stream function $\psi$ is then evaluated at the four neighboring locations surrounding the particle position through the selected spatial interpolation scheme together with linear temporal interpolation between two successive Eulerian time levels. The corresponding velocity components are subsequently reconstructed using second-order centered finite-difference approximations,

\begin{equation}
u(\mathbf{x}_p,t)=
\frac{
\psi(x_p,y_p+\Delta y,t)
-
\psi(x_p,y_p-\Delta y,t)}
{2\Delta y},
\qquad
v(\mathbf{x}_p,t)=
-
\frac{
\psi(x_p+\Delta x,y_p,t)
-
\psi(x_p-\Delta x,y_p,t)}
{2\Delta x},
\label{eq:particle_velocity_reconstruction}
\end{equation}

\hspace{-0.52cm}\textcolor{black}{where} $\mathbf{x}_p=(x_p,y_p)$ denotes the particle position at the corresponding RK4 stage, while $\Delta x$ and $\Delta y$ denote the Eulerian grid spacings. Finally, the particle position is advanced to the next time level by combining the four stage velocities according to the classical RK4 formula,

\begin{equation}
\mathbf{x}_p^{\,n+1}
=
\mathbf{x}_p^{\,n}
+
\frac{\Delta t_p}{6}
\left(
\mathbf{u}_1
+
2\mathbf{u}_2
+
2\mathbf{u}_3
+
\mathbf{u}_4
\right),
\label{eq:rk4_update}
\end{equation}

\hspace{-0.52cm}\textcolor{black}{where} $\mathbf{u}_1$, $\mathbf{u}_2$, $\mathbf{u}_3$, and $\mathbf{u}_4$ denote the particle velocities evaluated at the four RK4 stages.

The framework implements three spatial interpolation schemes, namely bilinear, bicubic Catmull--Rom, and bicubic B-spline interpolation. As demonstrated by the interpolation verification presented in Section~5, all three schemes yield nearly identical transport statistics for the turbulent flows considered in this work. 
Consequently, bilinear interpolation is employed throughout the production simulations owing to its lower computational cost without any observable loss of accuracy.

Periodic boundary conditions are employed during particle integration using the same periodic computational domain as the Eulerian fluid solver for visualization and animation. Whenever a particle crosses a domain boundary during the time integration, its position is remapped to the opposite side of the domain, thereby preserving the periodicity of the flow and enabling continuous particle trajectories for visualization and animation. For the computation of Lagrangian transport statistics, however, particle trajectories are unwrapped across the periodic boundaries so that the cumulative particle displacement is evaluated without artificial periodic remapping. Consequently, the mean-square displacement and diffusion coefficients are computed using open (unwrapped) particle trajectories.

The tracer-particle module is fully coupled to the Eulerian fluid solver and advanced using the same integration timestep. Particle positions are stored in Lagrangian form and evolved independently of the fixed Eulerian mesh. At each timestep, the Eulerian fluid fields are first advanced to the new time level, after which the updated flow fields are interpolated to the particle locations at each RK4 stage to advance the particle trajectories. This sequential integration strategy maintains synchronization between the Eulerian and Lagrangian components throughout the simulation while preserving one-way coupling from the fluid to the particles.

\subsection{Inertial-particle integration}

The inertial-particle equations introduced in Sec.~2 are integrated using the same classical fourth-order Runge--Kutta (RK4) framework employed for tracer particles. The numerical procedures for RK4 time integration, spatial and temporal interpolation, Eulerian--Lagrangian coupling, and synchronization with the Eulerian fluid solver remain identical to those described in Sec.~3.3. The principal difference is that inertial particles require the simultaneous evolution of both the particle position and particle velocity.

At each RK4 stage, intermediate particle positions and particle velocities are first predicted from the values obtained at the preceding stage. The fluid velocity is then interpolated directly from the Eulerian velocity field to the corresponding intermediate particle position using bilinear spatial interpolation together with linear temporal interpolation between successive Eulerian time levels. Unlike the tracer-particle solver, no stream-function reconstruction is required, since the particle acceleration depends directly on the local carrier-flow velocity. The interpolated fluid velocity is subsequently used to evaluate the Stokes-drag acceleration,

\begin{equation}
\mathbf{a}_p
=
\frac{\mathbf{u}(\mathbf{x}_p,t)-\mathbf{v}_p}{\mathrm{St}},
\label{eq:drag_rk4}
\end{equation}

where $\mathbf{a}_p$ is the particle acceleration, $\mathbf{x}_p=(x_p,y_p)$ denotes the particle position at the corresponding RK4 stage, $\mathbf{u}(\mathbf{x}_p,t)$ is the carrier-flow velocity interpolated from the Eulerian velocity field to the particle position, $\mathbf{v}_p$ is the particle velocity, and \textcolor{black}{$\mathrm{St}$ is the Stokes number.}

The above procedure is repeated at each of the four RK4 stages to evaluate the corresponding particle velocities and drag accelerations. The particle position and particle velocity are then advanced simultaneously using the classical RK4 weighted combination,

\begin{equation}
\mathbf{x}_p^{\,n+1}
=
\mathbf{x}_p^{\,n}
+
\frac{\Delta t_p}{6}
\left(
\mathbf{v}_1
+
2\mathbf{v}_2
+
2\mathbf{v}_3
+
\mathbf{v}_4
\right),
\qquad
\mathbf{v}_p^{\,n+1}
=
\mathbf{v}_p^{\,n}
+
\frac{\Delta t_p}{6}
\left(
\mathbf{a}_1
+
2\mathbf{a}_2
+
2\mathbf{a}_3
+
\mathbf{a}_4
\right).
\label{eq:rk4_inertial}
\end{equation}

\hspace{-0.52cm}\textcolor{black}{where} $\mathbf{x}_p^{\,n}$ and $\mathbf{v}_p^{\,n}$ denote the particle position and particle velocity at time level $t_n$, $\mathbf{x}_p^{\,n+1}$ and $\mathbf{v}_p^{\,n+1}$ denote the corresponding quantities at the next time level, $\Delta t_p$ is the particle-integration time step, and $\mathbf{v}_i$ and $\mathbf{a}_i$ $(i=1,\ldots,4)$ represent the particle velocity and particle acceleration evaluated at the four RK4 stages, respectively. The updated particle position and particle velocity are subsequently assigned as the current particle state and used as the initial conditions for the next timestep.

Based on the tracer-particle interpolation verification presented in Section~5, bilinear interpolation is adopted for all production inertial-particle simulations owing to its computational efficiency and demonstrated accuracy. Periodic boundary conditions are employed during particle integration using the same periodic computational domain as the Eulerian fluid solver for visualization and animation.  While wrapped particle trajectories are used for visualization and animation, unwrapped trajectories are retained for computing inertial-particle transport statistics, including the mean-square displacement and diffusion coefficients. As in the tracer-particle implementation, the present framework adopts a one-way coupling approximation, in which the carrier flow influences the particle motion while particle feedback on the fluid is neglected.

In the limit $\mathrm{St}\rightarrow0$, the particle velocity relaxes instantaneously to the local fluid velocity, thereby recovering the passive-tracer equations described in Sec.~3.3.

\subsection{Interpolation schemes}

Since Lagrangian particle positions generally do not coincide with Eulerian grid points, the fluid quantities required by the particle integrators are obtained through interpolation from the Eulerian mesh to the instantaneous particle locations. Three spatial interpolation schemes have been implemented and tested in the present Eulerian--Lagrangian framework: bilinear interpolation, bicubic Hermite (Catmull--Rom) interpolation, and bicubic spline interpolation. In all cases, linear interpolation in time is employed between two consecutive fluid time levels to provide the intermediate field values required by the RK4 particle integrator. Among these schemes, bilinear interpolation is adopted for all production simulations reported in this work owing to its computational efficiency and the interpolation verification presented in Sec.~5, whereas the bicubic Catmull--Rom and bicubic B-spline schemes are employed only for verification and convergence studies.

\subsubsection*{Bilinear interpolation}

For a scalar field \(f(x,y)\), representing either the stream function or a velocity component, let the particle position \((x_p,y_p)\) lie inside the grid cell bounded by

\begin{equation}
\xi=
\frac{x_p-x_i}{x_{i+1}-x_i},
\qquad
\eta=
\frac{y_p-y_j}{y_{j+1}-y_j},
\label{eq:bilinear_local}
\end{equation}

the bilinear interpolation is given by

\begin{equation}
\begin{aligned}
f(x_p,y_p) ={}&
(1-\xi)(1-\eta)\,f_{i,j}
+\xi(1-\eta)\,f_{i+1,j} 
+(1-\xi)\eta\,f_{i,j+1}
+\xi\eta\,f_{i+1,j+1}.
\end{aligned}
\label{eq:bilinear_interp}
\end{equation}

This interpolation procedure is applied independently to each flow variable required by the particle solver. For tracer-particle simulations, the stream function is interpolated and subsequently used to reconstruct the velocity components, whereas for inertial-particle simulations the velocity components are interpolated directly.

\subsubsection*{Bicubic Catmull-Rom interpolation}

The second interpolation option employs a bicubic Catmull--Rom interpolation scheme \textcolor{black}{[\cite{Li2016}]} based on the cubic Hermite formulation. Unlike bilinear interpolation, which utilizes only the four grid points surrounding the particle location, the Catmull--Rom method employs a local $4\times4$ stencil consisting of sixteen neighboring grid points, thereby providing a smoother and higher-order reconstruction of the Eulerian field.

Let the particle position $(x_p,y_p)$ lie within the grid cell bounded by $(x_i,y_j)$ and $(x_{i+1},y_{j+1})$. The local coordinates within the cell are defined as

\begin{equation}
t=\frac{x_p-x_i}{x_{i+1}-x_i},
\qquad
u=\frac{y_p-y_j}{y_{j+1}-y_j},
\label{eq:cr_local}
\end{equation}

where $t,u\in[0,1]$ denote the normalized distances from the lower-left corner of the cell.


\textcolor{black}{The interpolated field value is obtained through the tensor-product reconstruction}

\begin{equation}
f(x_p,y_p) = \sum_{m=1}^{4}
\sum_{n=1}^{4}
W_m(t)\,
W_n(u)\,
f_{i+m-2,j+n-2},
\label{eq:cr_interp}
\end{equation}

\textcolor{black}{where $f_{i+m-2,j+n-2}$ denotes the field values on the surrounding $4\times4$ stencil. The interpolation therefore combines information from sixteen neighboring grid points, in contrast to the four-point stencil employed by bilinear interpolation. The weighting functions \(W_m\) and \(W_n\) are the cubic Catmull--Rom basis functions, given by}

\begin{equation}
\begin{aligned}
W_1(s)&=-\frac{1}{2}s+s^2-\frac{1}{2}s^3,\quad
W_2(s)=1-\frac{5}{2}s^2+\frac{3}{2}s^3,\quad
W_3(s)=\frac{1}{2}s+2s^2-\frac{3}{2}s^3,\quad
W_4(s)=-\frac{1}{2}s^2+\frac{1}{2}s^3.
\end{aligned}
\label{eq:cr_kernel}
\end{equation}

\textcolor{black}{where $s=t$ for interpolation along the $x$ direction and $s=u$ for interpolation along the $y$ direction.}

The algorithm proceeds as follows. First, the computational cell containing the particle position is identified. Second, the local coordinates $(t,u)$ are computed using Eq.~(\ref{eq:cr_local}). Third, the Catmull--Rom kernel weights are evaluated from Eq.~(\ref{eq:cr_kernel}). Fourth, the surrounding $4\times4$ stencil is assembled and the interpolated value is computed using Eq.~(\ref{eq:cr_interp}). Near domain boundaries, where a complete $4\times4$ stencil is unavailable, the implementation automatically reverts to bilinear interpolation. This ensures numerical robustness while avoiding extrapolation beyond the available grid data.


The Catmull--Rom formulation provides a smooth higher-order reconstruction of the Eulerian field while remaining entirely local through the use of a $4\times4$ interpolation stencil.

\subsubsection*{Cubic B-spline interpolation}

In addition to bilinear and Catmull--Rom interpolation, a bicubic B-spline interpolation scheme\textcolor {black}{[\cite{deBoor2001}]} has also been implemented and tested within the present Eulerian--Lagrangian framework. Similar to the Catmull--Rom method, the bicubic B-spline employs a $4\times4$ stencil surrounding the particle position and performs interpolation using tensor-product cubic basis functions. However, unlike Catmull--Rom interpolation, which is an interpolating spline that passes exactly through the neighboring data points, the B-spline formulation is an approximating spline that produces a smoother reconstruction of the underlying field.

Let the particle position $(x_p,y_p)$ lie inside the computational cell bounded by $(x_i,y_j)$ and $(x_{i+1},y_{j+1})$. The normalized local coordinates are defined as

\begin{equation}
t_x=
\frac{x_p-x_i}{x_{i+1}-x_i},
\qquad
t_y=
\frac{y_p-y_j}{y_{j+1}-y_j},
\label{eq:bspline_local}
\end{equation}

where $0\le t_x,t_y\le1$.

\textcolor{black}{The interpolated field value is obtained through the tensor-product reconstruction}

\begin{equation}
f(x_p,y_p)=
\sum_{m=1}^{4}
\sum_{n=1}^{4}
B_m(t_x)\,
B_n(t_y)\,
f_{i+m-2,j+n-2},
\label{eq:bspline_interp}
\end{equation}

\textcolor{black}{where \(f_{i+m-2,j+n-2}\) denotes the field values on the surrounding \(4\times4\) stencil. Consequently, the interpolation utilizes information from sixteen neighboring grid points and yields a smooth approximation of the underlying field. The weighting functions \(B_m\) and \(B_n\) are the cubic B-spline basis functions, given by,}

\begin{equation}
\begin{aligned}
B_1(s)&=\frac{(1-s)^3}{6}, \qquad
B_2(s)=\frac{3s^3-6s^2+4}{6},\
B_3(s)&=\frac{-3s^3+3s^2+3s+1}{6}, \qquad
B_4(s)=\frac{s^3}{6},
\end{aligned}
\label{eq:bspline_kernel}
\end{equation}

\textcolor{black}{where $s=t_x$ for interpolation along the $x$ direction and $s=t_y$ for interpolation along the $y$ direction.}

The algorithm proceeds as follows. First, the computational cell containing the particle position is identified. Second, the local coordinates $(t_x,t_y)$ are computed using Eq.~(\ref{eq:bspline_local}). Third, the cubic B-spline weights are evaluated from Eq.~(\ref{eq:bspline_kernel}). Fourth, the surrounding $4\times4$ stencil is assembled and the interpolated field value is computed using Eq.~(\ref{eq:bspline_interp}). Near domain boundaries, where a complete $4\times4$ stencil is unavailable, the implementation reverts to bilinear interpolation to maintain numerical stability.


 Owing to its smooth cubic basis functions, the B-spline formulation provides a smooth approximation of the underlying field while retaining the locality of the $4\times4$ interpolation stencil.


\subsubsection*{Temporal interpolation}

The spatially interpolated field values at the intermediate RK4 stages are obtained by linear interpolation between two successive Eulerian time levels. If $f^n(x_p,y_p)$ and $f^{n+1}(x_p,y_p)$ denote the spatially interpolated field values at times $t^n$ and $t^{n+1}$, respectively, then

\begin{equation}
f(x_p,y_p,t)
=
(1-\theta)f^n(x_p,y_p)
+
\theta f^{n+1}(x_p,y_p),
\label{eq:temporal_interp}
\end{equation}

\hspace{-0.52cm}\textcolor{black}{where}

\begin{equation}
\theta=
\frac{t-t^n}{t^{n+1}-t^n},
\end{equation}

\hspace{-0.52cm}\textcolor{black}{with} $0\le\theta\le1$. This temporal interpolation procedure is employed together with each of the implemented spatial interpolation schemes to provide the intermediate field values required at the RK4 stages of the particle integrator.

The spatial and temporal interpolation procedures described above constitute the interpolation module employed throughout the present Eulerian--Lagrangian framework. Their influence on the computed Lagrangian transport statistics is assessed in Sec.~5.

\subsection{Coupled Eulerian--Lagrangian algorithm}

\textcolor{black}{The present framework adopts a one-way coupled Eulerian--Lagrangian strategy, in which the fluid influences the particle motion through the local velocity field, while the particles do not exert any feedback on the fluid. The incompressible Navier--Stokes equations are solved on a fixed Eulerian mesh, while particle trajectories are integrated in a Lagrangian reference frame.} At each global timestep, the Eulerian fluid solver first advances the spectral vorticity field using the pseudo-spectral second-order Adams--Bashforth scheme. The updated stream-function and velocity fields are subsequently reconstructed from
the spectral solution.

The particle solver is then advanced over the same timestep using the
classical fourth-order Runge--Kutta method. At each RK4 stage, the
required carrier-flow quantities are evaluated by combining spatial
interpolation with linear temporal interpolation between two successive
Eulerian time levels. For tracer particles, the stream function is
interpolated to the neighboring locations surrounding the intermediate
particle position, and the particle velocity is reconstructed using
centered finite differences. For inertial particles, the Eulerian
velocity field is interpolated directly to the intermediate particle
position and used to evaluate the Stokes-drag acceleration.

Intermediate particle positions are mapped periodically onto the
computational domain before the Eulerian fields are evaluated. After
completion of the RK4 update, the new particle state is assigned as the
current state for the next timestep. Wrapped particle coordinates are
used for periodic field evaluation and visualization, whereas unwrapped
trajectories are retained for the computation of transport statistics.
Because particle feedback on the fluid is neglected, information is
transferred only from the Eulerian fluid solver to the Lagrangian
particle solver, and the coupling remains one-way throughout the
simulation. The complete update sequence is summarized in
Algorithm~\ref{alg:coupled_EL}.

\begin{algorithm}[t]
\caption{Coupled Eulerian--Lagrangian update over one global timestep}
\label{alg:coupled_EL}
\begin{algorithmic}[1]

\For{$n=0,1,\ldots,N_t-1$}

    \State Evaluate the spectral right-hand side of the vorticity equation

    \State Advance $\hat{\omega}^{\,n}$ to $\hat{\omega}^{\,n+1}$
    using the pseudo-spectral AB2 fluid solver

    \State Reconstruct $\psi^{n+1}$ and $\mathbf{u}^{n+1}$
    from $\hat{\omega}^{\,n+1}$

    \State Retain the Eulerian fields at $t_n$ and $t_{n+1}$
    for temporal interpolation

    \For{each particle $p=1,2,\ldots,N_p$}

        \For{each RK4 stage $s=1,2,3,4$}

            \State Construct the intermediate particle state

            \State Map the intermediate particle position periodically
            onto the computational domain

            \If{tracer particle}

                \State Interpolate $\psi^n$ and $\psi^{n+1}$
                in space and time

                \State Reconstruct the stage velocity using
                centered finite differences

            \ElsIf{inertial particle}

                \State Interpolate $\mathbf{u}^n$ and
                $\mathbf{u}^{n+1}$ in space and time

                \State Evaluate the Stokes-drag acceleration
                $\mathbf{a}_p=
                [\mathbf{u}(\mathbf{x}_p,t)-\mathbf{v}_p]/\tau_p$

            \EndIf

        \EndFor

        \State Apply the classical RK4 update to the particle state

        \State Store wrapped coordinates for visualization and retain
        unwrapped coordinates for transport diagnostics

    \EndFor

    \State Assign the updated Eulerian and Lagrangian fields as the
    current states for the next timestep

\EndFor

\end{algorithmic}
\end{algorithm}

\section{Code Architecture}\label{}

\textcolor{black}{This section describes the software organization and implementation of the coupled Eulerian--Lagrangian framework. We first present the overall architecture of the code, highlighting the major computational modules and the data flow between the Eulerian and Lagrangian components. The parallel implementation is then described, including the GPU acceleration strategy, OpenACC-based parallelization, and the integration of the NVIDIA cuFFT library within the computational framework.}

\subsection{Overall software structure}

The coupled Eulerian--Lagrangian code is organized as a logically
modular computational framework comprising an Eulerian fluid solver,
FFT interface, interpolation routines, Lagrangian particle integrators,
and diagnostic and output components. The overall execution is
controlled by a main driver program, which initializes the fluid and
particle variables, advances the coupled system in time, and manages
data storage and diagnostic calculations.

The Eulerian component is based on the GPU-accelerated GHD2D solver.
It advances the two-dimensional incompressible Navier--Stokes equations
in vorticity--streamfunction form using the Fourier pseudospectral and
second-order Adams--Bashforth procedures described in Sec.~3.1. The
spectral operations and Fourier transforms are performed through a
dedicated interface to the NVIDIA cuFFT library. At every fluid
timestep, this component provides the updated vorticity, stream-function,
and velocity fields on the fixed Eulerian mesh.

The interpolation routines provide the data-transfer layer between the
Eulerian and Lagrangian descriptions. They evaluate the required fluid
quantities at arbitrary particle locations using the spatial and
temporal interpolation procedures described in Sec.~3.4. In the present
implementation, the interpolation procedures are contained within the
main Fortran program and are called directly during the RK4 particle
stages.

The Lagrangian component contains separate numerical procedures for
tracer and inertial particles. The tracer-particle procedure advances
particle positions using the locally evaluated fluid velocity, whereas
the inertial-particle procedure advances both particle positions and
particle velocities under the Stokes-drag model. Both procedures employ
classical fourth-order Runge--Kutta integration and use the common
interpolation infrastructure.

The diagnostic and output component computes fluid quantities such as
kinetic energy and low-order field moments and stores Eulerian fields
and particle trajectories at prescribed output intervals. Particle
coordinates may be stored in wrapped form for visualization and
animation or in unwrapped form for subsequent transport analysis.

\begin{figure*}
    \centering
    \includegraphics[
        width=\textwidth,
        keepaspectratio
    ]{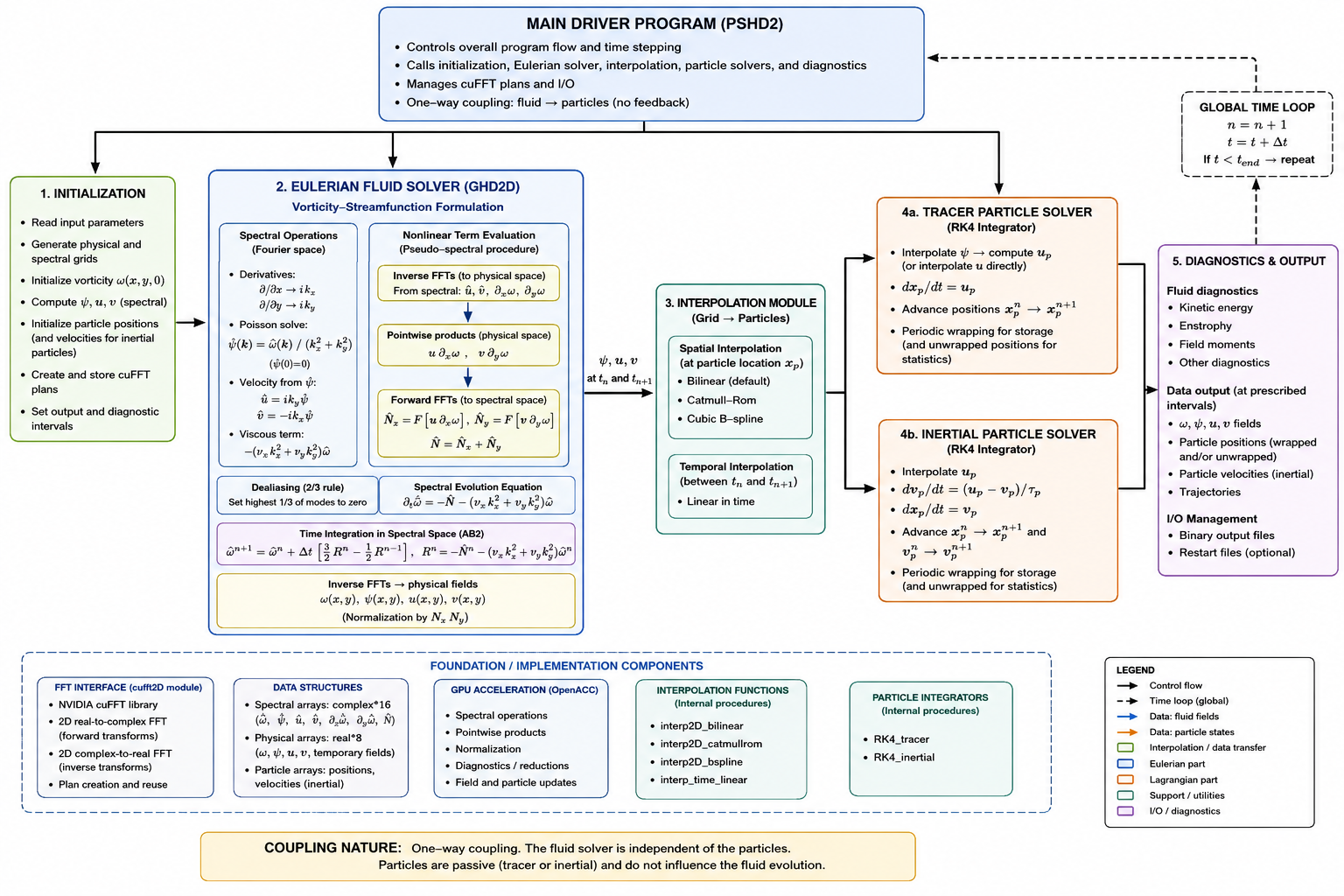}
    \caption{
    Logical software architecture and data flow of the coupled
    Eulerian--Lagrangian framework. The main driver controls
    initialization, Eulerian fluid advancement, interpolation,
    particle integration, diagnostics, and output. The GPU-accelerated
    GHD2D solver advances the vorticity--streamfunction equations using
    Fourier pseudospectral discretization and second-order
    Adams--Bashforth time integration. The updated stream-function and
    velocity fields are transferred to the tracer- and inertial-particle
    RK4 integrators through spatial and temporal interpolation. Wrapped
    particle coordinates are retained for visualization, while
    unwrapped trajectories are used for transport statistics. The
    absence of feedback from the particle solvers to the Eulerian fluid
    solver indicates the one-way coupling adopted in the present
    implementation. \textcolor{black}{The schematic was prepared with the assistance of ChatGPT (OpenAI) based on the software structure and workflow specified by the authors.}
    }
    \label{fig:software_architecture}
\end{figure*}

Although the current implementation is assembled within a common
driver program, the fluid, interpolation, particle-integration, and
diagnostic operations are separated logically according to their
numerical functions. This organization permits the individual
components to be verified independently and facilitates the future
addition of alternative particle models, interpolation procedures, or
fluid solvers. A schematic representation of the principal software
components and their data exchange is shown in
Fig.~\ref{fig:software_architecture}.

\subsection{Parallel implementation}

The Eulerian component of the present framework is accelerated on a single NVIDIA GPU using the OpenACC programming model together with the NVIDIA cuFFT library. Since the computational cost of the coupled Eulerian--Lagrangian solver is dominated by the pseudospectral fluid
solver, GPU acceleration is applied to the computationally intensive Eulerian operations, including Fourier transforms, spectral differentiation, evaluation of nonlinear terms, dealiasing, velocity reconstruction, time integration, and diagnostic calculations.

The Fourier transforms required by the pseudospectral formulation are performed through custom Fortran--C interfaces to the NVIDIA cuFFT library. Separate double-precision real-to-complex and complex-to-real transform plans are created once during the initialization stage and subsequently reused throughout the simulation,
thereby avoiding repeated FFT planning overhead. Communication between OpenACC-managed device memory and the cuFFT routines is achieved through the \texttt{host\_data use\_device} construct, allowing the Fourier transforms to operate directly on GPU-resident arrays without intermediate host--device copies.

Parallelism is exposed through OpenACC directives applied to the
dominant two-dimensional computational kernels. Most Eulerian
operations are parallelized using
\texttt{parallel loop collapse(2)} directives, thereby exposing both
spatial dimensions simultaneously to the GPU and improving thread
occupancy. These kernels include the evaluation of spectral operators,
solution of the Poisson equation, reconstruction of the velocity field,
computation of vorticity gradients, evaluation of the nonlinear
advection term, dealiasing operations, Adams--Bashforth updates, and the
normalization of inverse-transformed fields. Global diagnostic
quantities, including kinetic energy and low-order field moments, are
computed efficiently using OpenACC reduction operations.

During the fluid integration, the principal Eulerian field arrays remain
resident on the GPU within OpenACC data regions whenever possible,
thereby minimizing host--device communication. Successive spectral
operations, nonlinear evaluations, Fourier transforms, and time
integration are therefore executed almost entirely on the device.
Synchronization with host memory is performed only when required for
particle integration, diagnostic output, or data storage.

The present implementation adopts a heterogeneous execution strategy for
the coupled Eulerian--Lagrangian framework. At each global timestep, the
GPU-accelerated Eulerian solver first advances the fluid equations and
reconstructs the updated stream-function and velocity fields. These
fields are then transferred to the host, where the Lagrangian particle
integrators perform the RK4 trajectory integration using the spatial and
temporal interpolation procedures described in Sec.~3.4. Because the
computational cost is dominated by the pseudospectral fluid solver and
its associated FFT operations, the additional overhead introduced by the
CPU-based particle integration remains comparatively small for the
particle populations considered in the present work.

The logical separation of the Eulerian solver, interpolation routines,
particle integrators, and diagnostic components facilitates future
extensions of the framework. In particular, alternative interpolation
schemes, additional particle models, or GPU acceleration of the
Lagrangian particle modules can be incorporated without modification of
the underlying Eulerian fluid solver.

\section{Verification Tests}\label{}

\textcolor{black}{This section presents a comprehensive verification of the individual components of the coupled Eulerian--Lagrangian framework. The Eulerian fluid solver, tracer-particle module, inertial-particle module, interpolation procedures, and statistical convergence of the Lagrangian transport calculations are each validated using analytical solutions, established benchmark problems, and numerical convergence studies. These verification tests collectively establish the accuracy and robustness of the numerical framework prior to its application to turbulent flow simulations.}

\subsection{Fluid solver verification}\label{}

To verify the Eulerian fluid solver, we consider the classical Kelvin–Helmholtz instability (KHI) in an incompressible shear flow. The KHI provides a standard benchmark for assessing the ability of numerical solvers to accurately capture the growth of hydrodynamic instabilities driven by velocity shear. For the broken-jet configuration considered here, the viscous instability growth rate is given analytically by [\cite{D1961}],

  \begin{equation}
\gamma = \frac{k_x U_0}{3} \left[ \sqrt{3} - 2\frac{k_x}{R_E} - 2 \sqrt{\left( \frac{k_x}{R_E} \right)^2 + 2\sqrt{3} \frac{k_x}{R_E} } \right] \label{eq:KHI}
\end{equation}

\hspace{-0.52cm}\textcolor{black}{where} $\gamma$ is the linear growth rate of the Kelvin--Helmholtz instability, while $k_x$, $U_0$, and $R_E$ denote the perturbation wavenumber, characteristic shear velocity, and Reynolds number, respectively.

\begin{figure*}
    \centering
    \begin{subfigure}[t]{0.49\textwidth}  
        \centering
        \includegraphics[width=\textwidth]{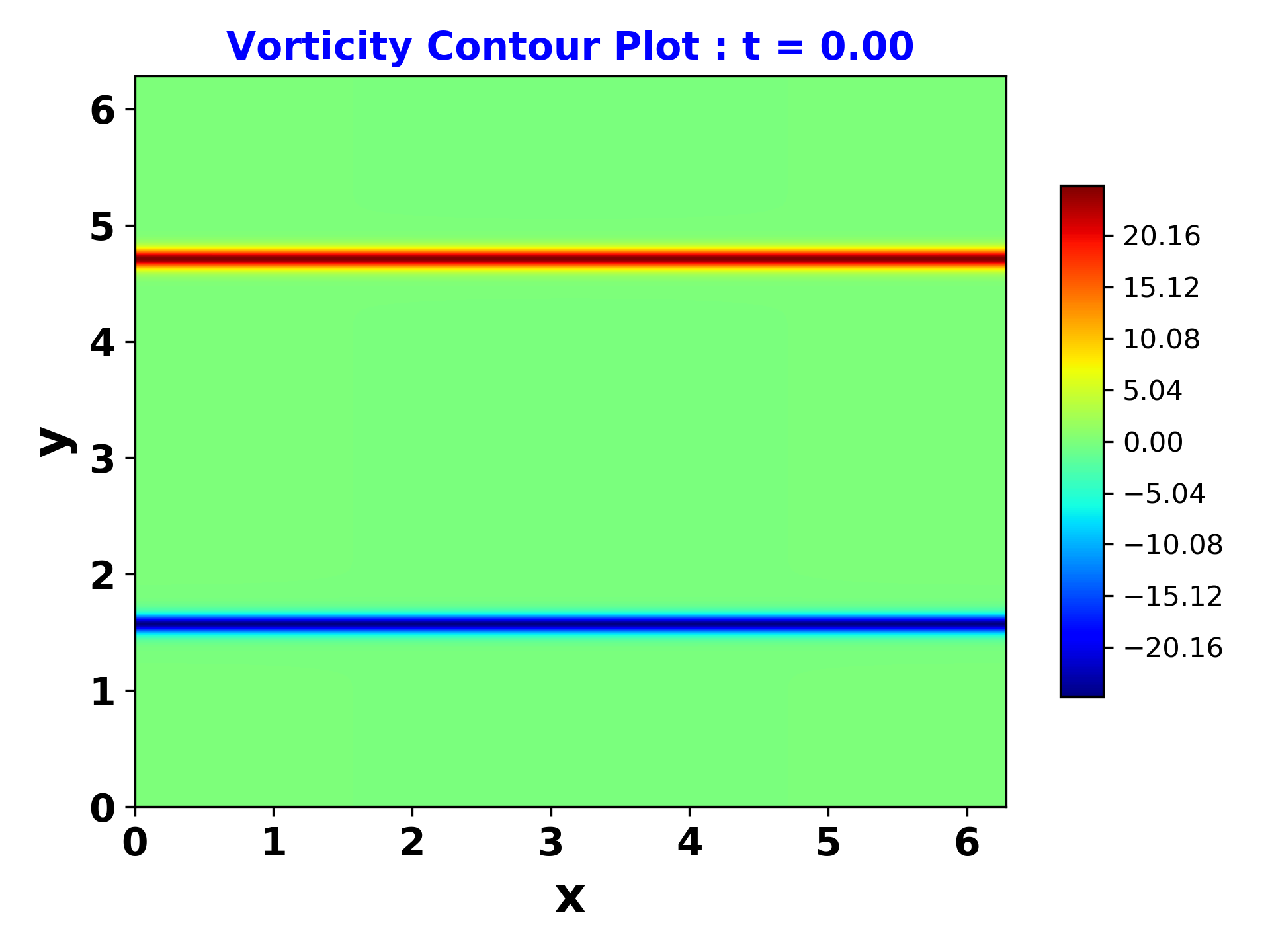}
        \caption{}
        \label{brokenjet_ic}
    \end{subfigure}
    \hfill
    \begin{subfigure}[t]{0.49\textwidth}  
        \centering
        \includegraphics[width=\textwidth]{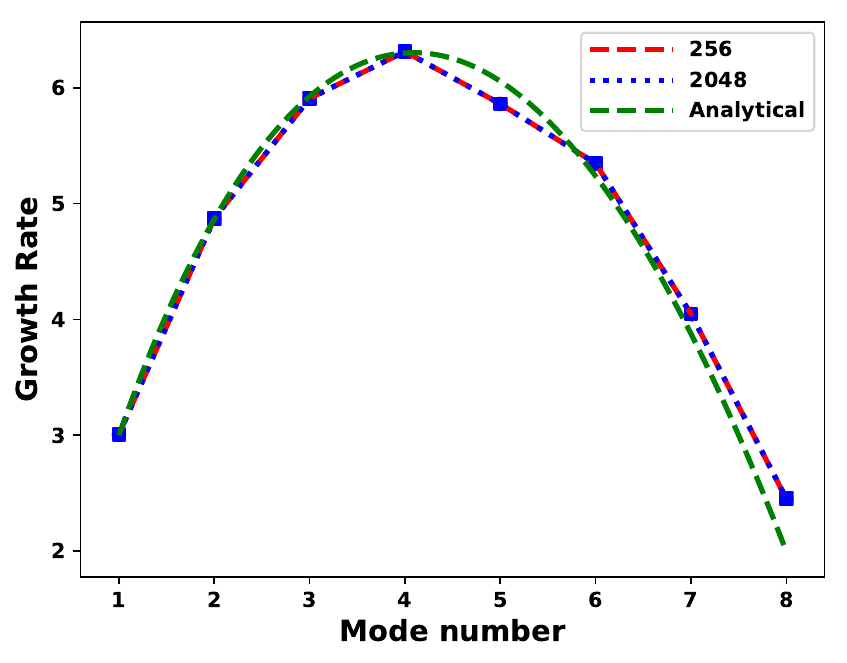}
        \caption{}
        \label{growth_analytical_fitting}
    \end{subfigure}
    \caption{(a) Initial condition: Two finite strips of fluid with alternating vorticity, flowing in opposite directions, forming discontinuous shear layers (commonly referred to as “oppositely directed broken jets”) to study Kelvin–Helmholtz instability. The Reynolds number of the flow is $R_E = 122$, and the peak vorticity is $\omega_0 = \pm 25$. (b) The growth rates of the Kelvin–Helmholtz instability in the oppositely directed broken jets problem, computed using the GHD2D solver for various mode numbers, are in excellent agreement with the analytical results of Drazin \textit{et al.} in the two-strip configuration. Simulations were performed at grid resolutions of $256^2$ and $2048^2$, and the results are found to be invariant with respect to resolution. 
    \label{brokenjet_benchmark}}
\end{figure*}

The initial condition consists of two oppositely directed vorticity strips forming a broken-jet profile, as shown in Fig.~\ref{brokenjet_ic}. The instability growth rate is obtained from the temporal evolution of the perturbation amplitude and compared with the analytical prediction. Figure~\ref{growth_analytical_fitting} shows the growth rate as a function of mode number for $R_E$=122. Excellent agreement is observed between the numerical and analytical results over the range of unstable modes considered. Furthermore, simulations performed at resolutions of $(256^2)$ and $(2048^2)$ yield nearly identical growth rates, demonstrating numerical convergence with respect to grid resolution.

The close agreement with the analytical solution confirms the accuracy of the GHD2D fluid solver and establishes its suitability for the coupled Eulerian–Lagrangian simulations presented in the following sections.

\subsection{ Tracer particle solver verification} 

\begin{figure*}
\centering

\begin{subfigure}{0.48\linewidth}
\centering
\includegraphics[width=\linewidth,height=0.22\textheight,keepaspectratio]{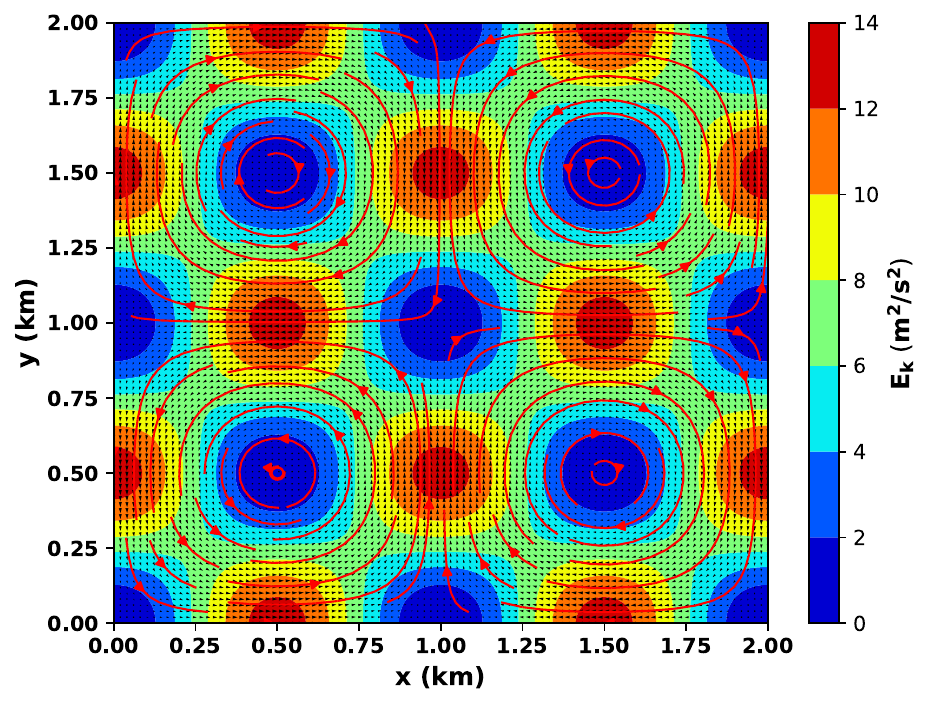}
\caption{}
\label{fig:kinv}
\end{subfigure}
\hfill
\begin{subfigure}{0.48\linewidth}
\centering
\includegraphics[width=\linewidth,height=0.22\textheight,keepaspectratio]{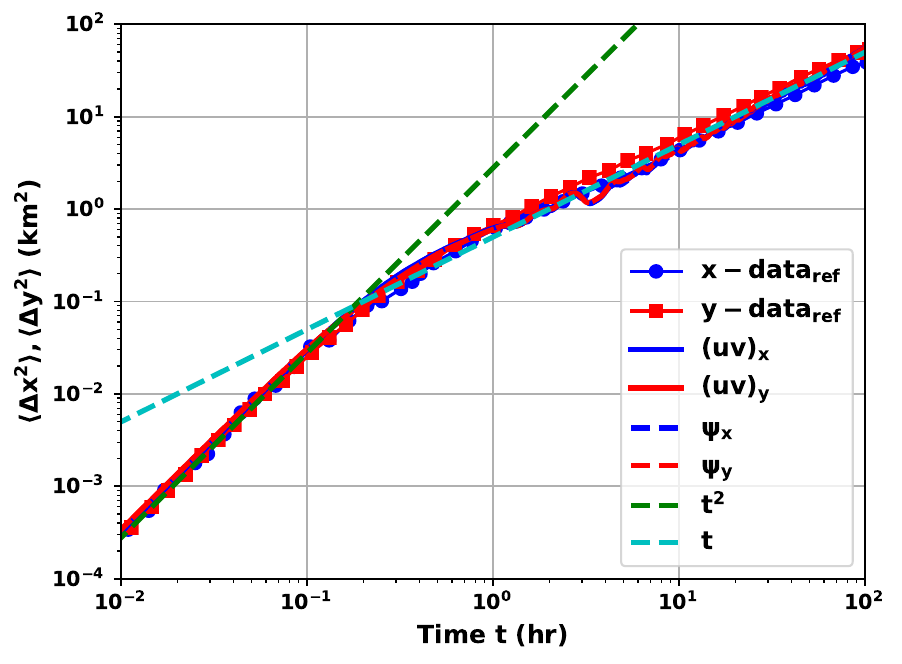}
\caption{}
\label{fig:kinp}
\end{subfigure}

\caption{(a) Kinematic chaotic-flow (2D oscillating Taylor-Green flow) benchmark of ~\cite{FL2022} showing the periodic lattice of non-stationary eddies used for tracer-particle verification. (b) Mean-square displacement of tracer particles obtained using analytical velocity evaluation, velocity-field interpolation, and stream-function interpolation. All approaches reproduce the ballistic-to-diffusive transition reported by [\cite{FL2022}]}. 
\label{fig:LagBM}
\end{figure*}

We verify the tracer-particle module using the benchmark problem of tracer transport in a two-dimensional time-periodic kinematic flow exhibiting chaotic advection (oscillating Taylor--Green flow) adopted from~[\cite{FL2022}]. The flow consists of a periodic lattice of non-stationary kinematic eddies (Fig.~\ref{fig:kinv}) and provides a convenient test case for assessing the accuracy of particle advection and interpolation procedures in a prescribed velocity field. The stream-function and velocity fields are given in Eq.~\eqref{eq:kinematic_flow}\textcolor{black}{~[\cite{FL2022}]}.

\begin{subequations}
\begin{align}
\psi &= \frac{\alpha}{k}
\sin\!\left(k[x-\epsilon\sin(\omega t)]\right)
\sin\!\left(k[y-\epsilon\sin(\omega t+\phi)]\right), \\
u &= \alpha
\sin\!\left(k[x-\epsilon\sin(\omega t)]\right)
\cos\!\left(k[y-\epsilon\sin(\omega t+\phi)]\right), \\
v &= -\alpha
\cos\!\left(k[x-\epsilon\sin(\omega t)]\right)
\sin\!\left(k[y-\epsilon\sin(\omega t+\phi)]\right).
\end{align}
\label{eq:kinematic_flow}
\end{subequations}

The reference configuration consists of a square domain of length \(L = 2~\mathrm{km}\) with spatial resolution \(\Delta x = 0.02~\mathrm{km}\). The corresponding spatial wavenumber of the imposed flow is \(k = 2\pi/L = \pi~\mathrm{km}^{-1}\), yielding an eddy size of \(L/2 = 1~\mathrm{km}\). The maximum flow speed is \(\alpha = 3.6~\mathrm{km\,hr^{-1}}\), while the oscillation parameters are chosen as \(\varepsilon/L = 0.3\), \(\omega \approx \alpha/(2\pi k^{-1}) = 1.8~\mathrm{hr^{-1}}\), and \(\phi = 0\), corresponding to a fully chaotic Lagrangian transport regime. An ensemble of 1000 passive tracer particles is evolved for 100 hours using a time step of \(\Delta t = 10^{-3}~\mathrm{hr}\).

The benchmark is particularly useful for validating the interpolation strategy employed in the Eulerian–Lagrangian coupling. Since the velocity field is analytically prescribed, particle trajectories and transport statistics obtained from interpolation-based approaches can be directly compared with those obtained from analytical evaluation of the flow field. We therefore compute tracer trajectories using three approaches: (i) direct evaluation of the analytical velocity field, (ii) interpolation of the velocity field, and (iii) interpolation of the stream-function field followed by evaluation of the corresponding velocity components. The latter approach corresponds to the procedure employed in the coupled Eulerian–Lagrangian solver.

Figure~\ref{fig:kinp} compares the resulting mean-square displacement obtained using the different approaches. Excellent agreement is observed among all methods throughout the simulation, demonstrating that the interpolation procedure accurately reproduces the underlying particle dynamics. In all cases, the transport exhibits an initial ballistic regime followed by a transition to normal diffusion at later times, consistent with the reference results of ~[\cite{FL2022}]. The close agreement between the analytical and interpolation-based solutions confirms the accuracy and robustness of the tracer-particle implementation and the interpolation strategy adopted in the present Eulerian–Lagrangian framework.

The agreement between the analytical and interpolation-based solutions demonstrates the accuracy of the tracer-particle integration scheme and the associated interpolation procedures. We emphasize that the particle solver advances Lagrangian trajectories solely from the local fluid velocity evaluated at the particle position and does not rely on assumptions specific to the underlying flow structure. Consequently, the validation presented here is independent of the dynamical origin of the velocity field and is equally applicable to kinematic, turbulent, or otherwise time-dependent flows. The present benchmark therefore establishes the numerical correctness of the tracer-particle module prior to its application to dynamically evolving turbulence simulations.

\subsection{Inertial particle solver verification} 

The inertial-particle module is verified using both analytical and numerical benchmark problems. The first benchmark considers the motion of a single inertial particle in a uniform velocity field, for which an exact analytical solution exists. The second benchmark examines inertial particle dynamics in a steady Taylor--Green vortex flow and compares the resulting trajectories with previously reported results by [\cite{Nath2024PRF}].

For the analytical benchmark, we consider a constant velocity field
$\mathbf{u}=(u_x,u_y)=(1,0)$.
In the absence of an intrinsic flow timescale, time is nondimensionalized using a reference timescale set to unity, such that the Stokes number is equivalent to the particle response time, $\mathrm{St}=\tau_p$.
For the initial conditions

\begin{equation}
x(0)=0.25,\qquad y(0)=0.25,\qquad
v_x(0)=0,\qquad v_y(0)=0,
\end{equation}

the analytical solution of Eq.~\eqref{eq:inertial_particle} is

\begin{equation}
v_x(t)=1-e^{-t/\tau_p},
\qquad
a_x(t)=\frac{1}{\tau_p}e^{-t/\tau_p},
\qquad
x(t)=x_0+t-\tau_p\left(1-e^{-t/\tau_p}\right).
\label{eq:uniform_solution}
\end{equation}

\begin{figure*}
\centering
\includegraphics[width=0.55\columnwidth]{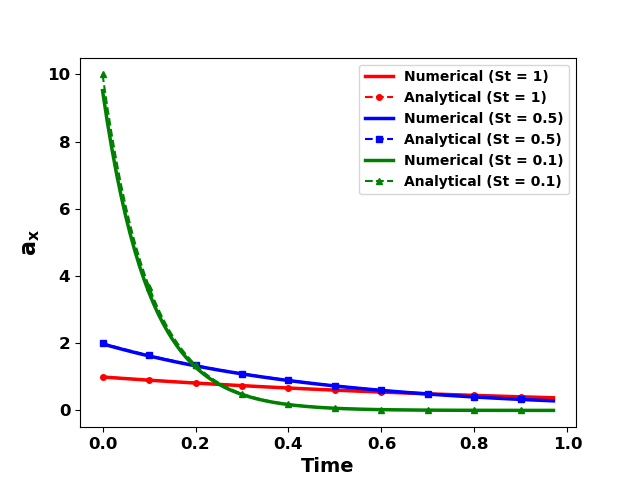}
\caption{Comparison of numerical and analytical particle acceleration $a_x(t)$ in a uniform flow for different Stokes numbers. The numerical results (solid lines) closely match the analytical solution (dashed lines), demonstrating accurate resolution of exponential relaxation dynamics.}
\label{fig:uniform_validation}
\end{figure*}

The particle velocity therefore relaxes exponentially toward the carrier-flow velocity, while the particle acceleration exhibits a corresponding exponential decay. Figure~\ref{fig:uniform_validation} compares the numerical and analytical particle accelerations for several Stokes numbers. Excellent agreement is observed for both the relaxation timescale and amplitude of the response. As an additional verification, the transit time required for a particle to move from (x=0.25) to (x=0.5) is computed numerically and compared with the analytical prediction (Table~\ref{tab:uniform_validation}). The numerical results are found to be in excellent agreement with the analytical solution over the full range of Stokes numbers considered.

\begin{table}[h!]
\centering
\caption{Time required for an inertial particle to move from $x=0.25$ to $x=0.5$ for different Stokes numbers ($\mathrm{St}$) in a uniform flow. The limit $\mathrm{St}=0$ corresponds to tracer motion.}
\begin{tabular}{|c|c|c|c|c|c|c|c|c|c|c|c|c|}
\hline
$\mathrm{St}$ & 0 & 0.01 & 0.1 & 0.2 & 0.3 & 0.4 & 0.5 & 0.6 & 0.7 & 0.8 & 0.9 & 1.0 \\ \hline
$t$ & 0.250 & 0.260 & 0.347 & 0.426 & 0.492 & 0.548 & 0.599 & 0.645 & 0.688 & 0.728 & 0.766 & 0.801 \\ \hline
\end{tabular}
\label{tab:uniform_validation}
\end{table}

Following the analytical validation, the inertial-particle module is further benchmarked using the steady Taylor--Green vortex flow studied by ~\cite{Nath2024PRF}. This benchmark is particularly useful because inertial particles exhibit a wide range of dynamical behaviors depending on their initial conditions and Stokes numbers, including periodic orbits, chaotic attractors, stable-point trapping, diffusive wandering, and ballistic transport. Using the same initial conditions and particle parameters reported in the reference study, the present implementation successfully reproduces all major classes of particle dynamics.

\begin{figure*}

\begin{minipage}{0.49\textwidth}
\centering
\includegraphics[width=\textwidth]{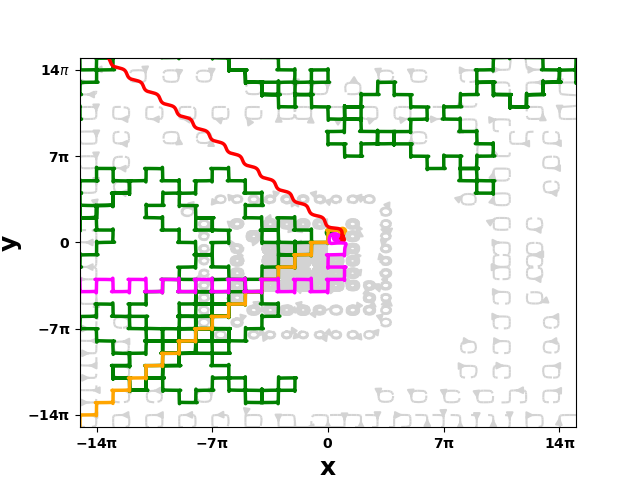}
\textbf{(a)}
\end{minipage}
\hfill
\begin{minipage}{0.49\textwidth}
\centering
\includegraphics[width=\textwidth]{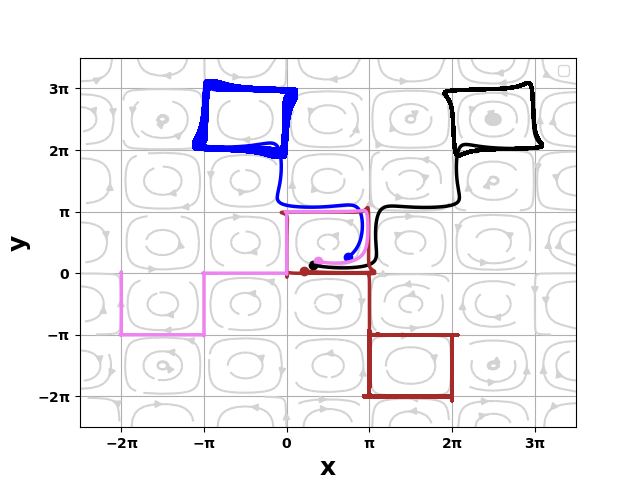}
\textbf{(b)}
\end{minipage}

\caption{ Representative particle dynamics. (a) Transport behaviors:
(i) $\textcolor{green}{(0.12,\,2.6)}$, $\mathrm{St}=1.12$ showing diffusive motion;
(ii) $\textcolor{orange}{(2.6,\,2.8)}$, $\mathrm{St}=0.95$ exhibiting diagonal square ballistic motion; (iii) $\textcolor{red}{(2.48,\,1.07)}$, $\mathrm{St}=1.83$ showing diagonal smooth ballistic motion; (iv) $\textcolor{magenta}{(1.5,\,1.5)}$, $\mathrm{St}=1.133$ exhibiting horizontal ballistic motion. (b) Bounded dynamics: (i) $(1.0,\,0.42)$ with $\mathrm{St}=1.12$ corresponding to a limit cycle; (ii) $\textcolor{blue}{(2.33,\,0.81)}$, $\mathrm{St}=1.37$ showing a chaotic attractor; (iii) $\textcolor{brown}{(0.664,\,0.11)}$, $\mathrm{St}=0.89$ trapped in a square limit cycle; (iv) $\textcolor{violet}{(1.2,\,0.6)}$, $\mathrm{St}=0.5$ trapped at the stable point (SP).
}
\label{fig:particle_dynamics_SS}
\end{figure*}

Representative trajectories are shown in Fig.~\ref{fig:particle_dynamics_SS}. The numerical solutions recover both bounded and unbounded particle motions, including limit cycles, chaotic attractors, trapped states, diffusive trajectories, and ballistic transport. The close agreement with previously reported results confirms the accuracy of the inertial-particle integration scheme in spatially varying velocity fields.

The analytical and Taylor--Green benchmarks together validate the inertial-particle module over a broad range of dynamical regimes. Since the particle trajectories are advanced solely through the local fluid velocity evaluated at the particle position, the verification presented here is independent of the specific flow configuration employed. The benchmarks therefore establish the numerical correctness and robustness of the inertial-particle solver prior to its application to more complex time-dependent and turbulent flows.

\subsection{Interpolation verification}

\begin{figure*}
\centering

\begin{subfigure}{0.48\linewidth}
\centering
\includegraphics[width=\linewidth,height=0.22\textheight,keepaspectratio]{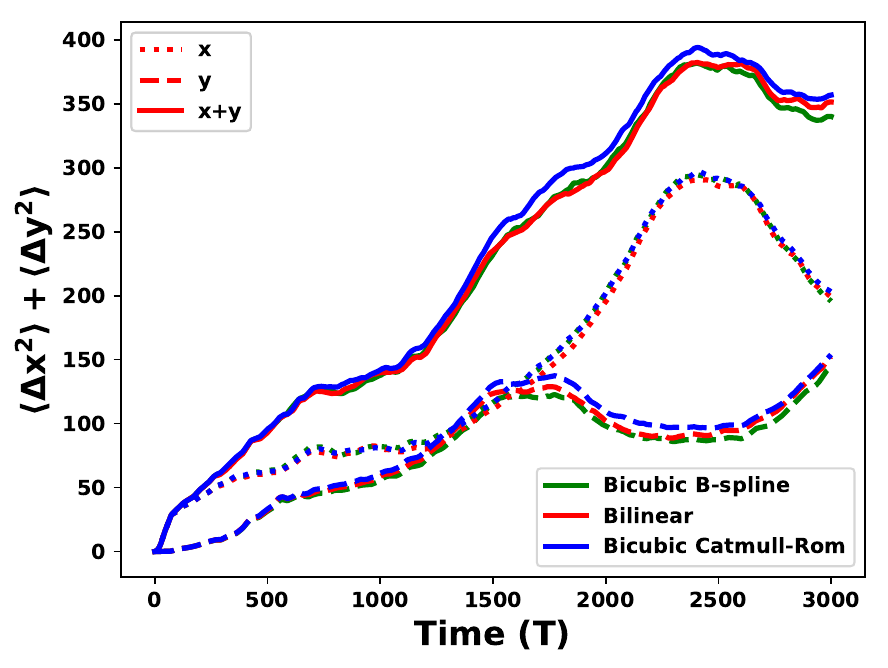}
\caption{\textbf{VPF = 12.5\% (4-strip configuration)}}
\label{fig:interp4}
\end{subfigure}
\hfill
\begin{subfigure}{0.48\linewidth}
\centering
\includegraphics[width=\linewidth,height=0.22\textheight,keepaspectratio]{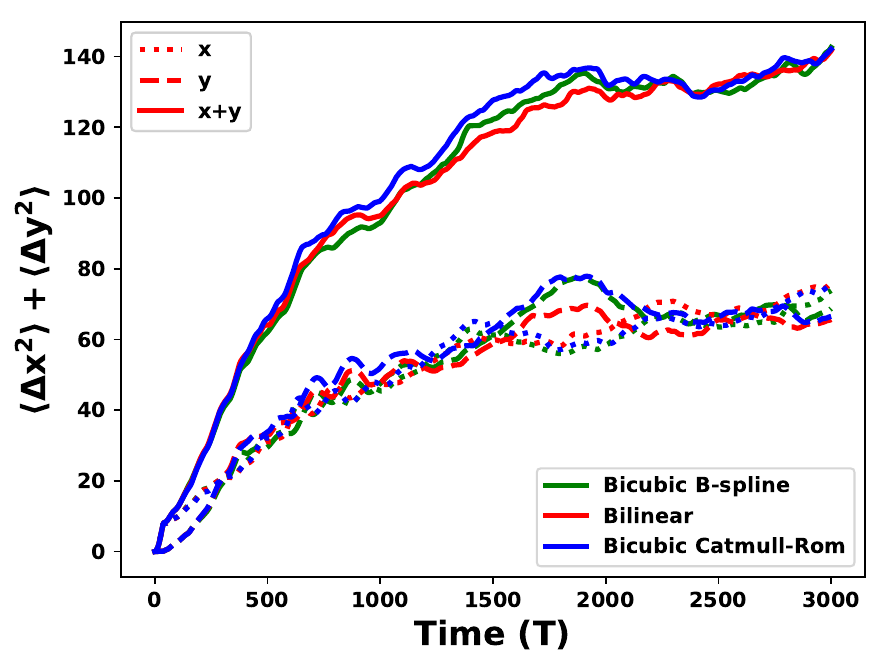}
\caption{\textbf{VPF = 50\% (16-strip configuration)}}
\label{fig:interp16}
\end{subfigure}

\caption{
Comparison of the mean-square displacement (MSD) obtained using the three interpolation schemes implemented in the present framework: \textbf{bilinear}, \textbf{bicubic Catmull--Rom}, and \textbf{bicubic B-spline} interpolation. Results are shown for two representative initial vorticity packing fractions (VPFs): (a) VPF = 12.5\% (4-strip configuration) and (b) VPF = 50\% (16-strip configuration). The three interpolation schemes produce nearly indistinguishable MSD curves throughout the simulation, demonstrating that the computed Lagrangian transport statistics are insensitive to the interpolation method. Consequently, the computationally efficient bilinear interpolation is employed in all production simulations.
}
\label{fig:interp_verification}
\end{figure*}

To evaluate the sensitivity of the computed transport statistics to the interpolation scheme, simulations were carried out using the three interpolation methods implemented in the present framework: bilinear, bicubic Catmull--Rom, and bicubic B-spline interpolation. As a representative test of the interpolation strategy in the coupled Eulerian--Lagrangian framework, we consider two decaying two-dimensional Navier--Stokes simulations corresponding to the 4-strip (VPF = 12.5\%) and 16-strip (VPF = 50\%) initial configurations introduced in Section~6. These cases span distinct turbulent flow regimes and therefore provide a stringent assessment of the interpolation strategy. Figure~\ref{fig:interp_verification} compares the resulting mean-square displacement (MSD) obtained using the three interpolation schemes.

The MSD obtained using the three interpolation schemes is found to be in excellent agreement for both representative VPF configurations throughout the simulation. Only negligible differences are observed, indicating that the computed particle dynamics are largely insensitive to the choice of interpolation scheme. Although the bicubic Catmull--Rom and bicubic B-spline methods employ higher-order spatial reconstruction, they do not yield any appreciable improvement in the computed transport statistics compared with the simpler bilinear interpolation. These results demonstrate that bilinear interpolation provides sufficient accuracy for the present Eulerian--Lagrangian framework while offering superior computational efficiency. Consequently, bilinear interpolation is adopted for all production simulations presented in this work.


\subsection{Particle-number convergence}

Having established the accuracy of the interpolation procedure, we next assess the statistical convergence of the computed Lagrangian transport quantities with respect to the number of tracer particles employed. To this end, particle-number convergence tests are performed using representative decaying two-dimensional Navier--Stokes simulations at the full numerical resolution of $(2048^2)$. Simulations are carried out with $N_p=100$, $500$, $1000$, and $2000$ tracer particles, and the resulting transport statistics are compared.

\begin{figure*}
     \centering
     \begin{subfigure}[h]{0.49\textwidth}
         \centering
         \includegraphics[width=\textwidth]{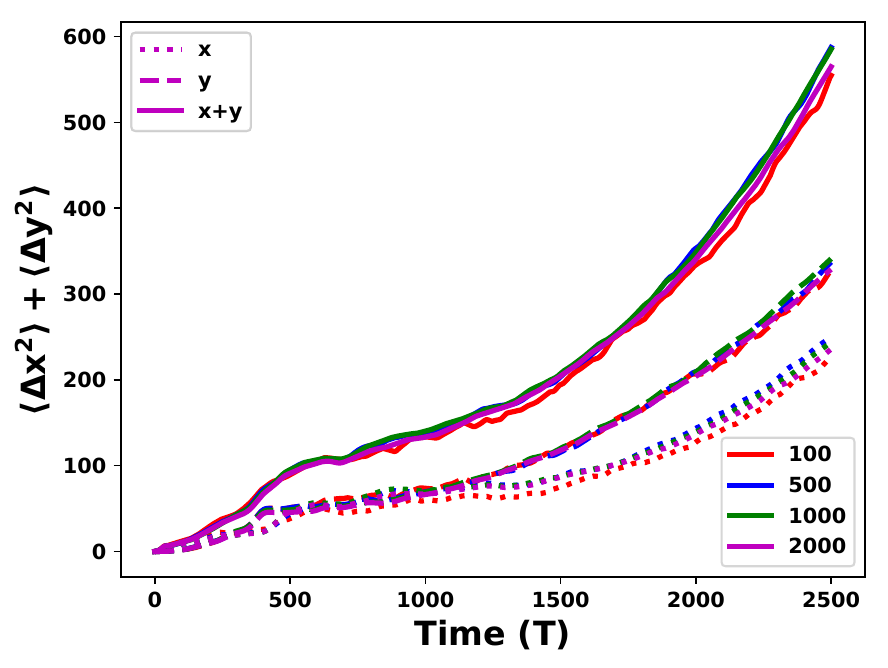}
        \caption{}
    \end{subfigure}
    \hfill
    \begin{subfigure}[h]{0.49\textwidth}
         \centering
         \includegraphics[width=\textwidth]{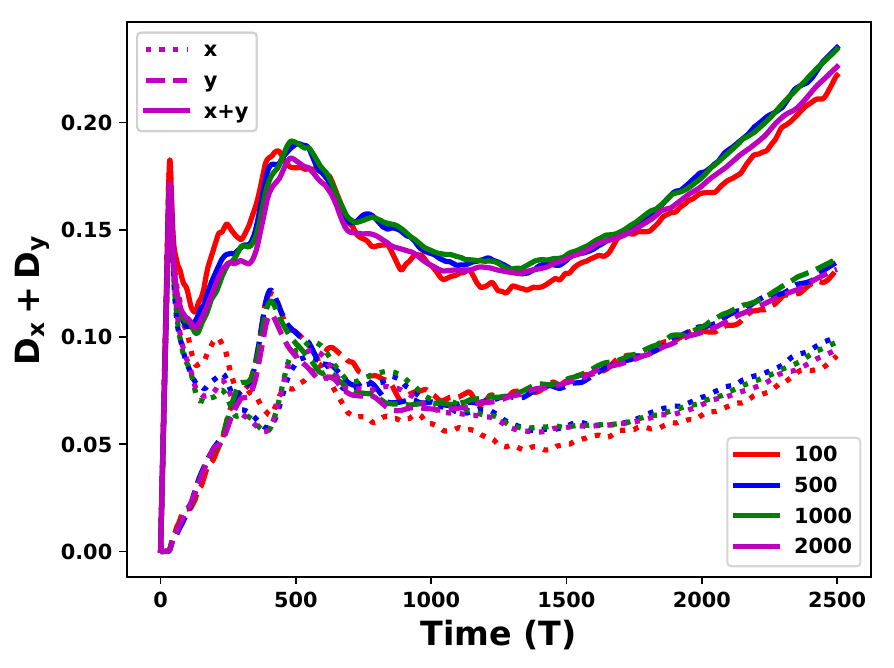}
        \caption{}
    \end{subfigure}
    \caption{{{ Particle-number convergence study performed using decaying two-dimensional turbulence simulations at ($2048^2$) resolution. (a) Mean-square displacement and (b) diffusion coefficients obtained using ($N_p$=100, 500, 1000, and 2000 tracer particles. The transport statistics are largely insensitive to particle number, indicating convergence of the Lagrangian transport measures.}}}
        \label{fig:particle_convergence1}

\end{figure*}

Figure~\ref{fig:particle_convergence1} shows the evolution of the mean-square displacement (MSD) and the corresponding diffusion coefficients obtained for different particle numbers. The  transport statistics are found to be nearly indistinguishable over the entire simulation interval, indicating weak sensitivity of the transport statistics to the particle ensemble size. In particular, the MSD curves obtained using ($N_p$ = 1000) and ($N_p$ = 2000) particles are virtually identical.

\begin{figure*}
     \centering
     \begin{subfigure}[h]{0.49\textwidth}
         \centering
         \includegraphics[width=\textwidth]{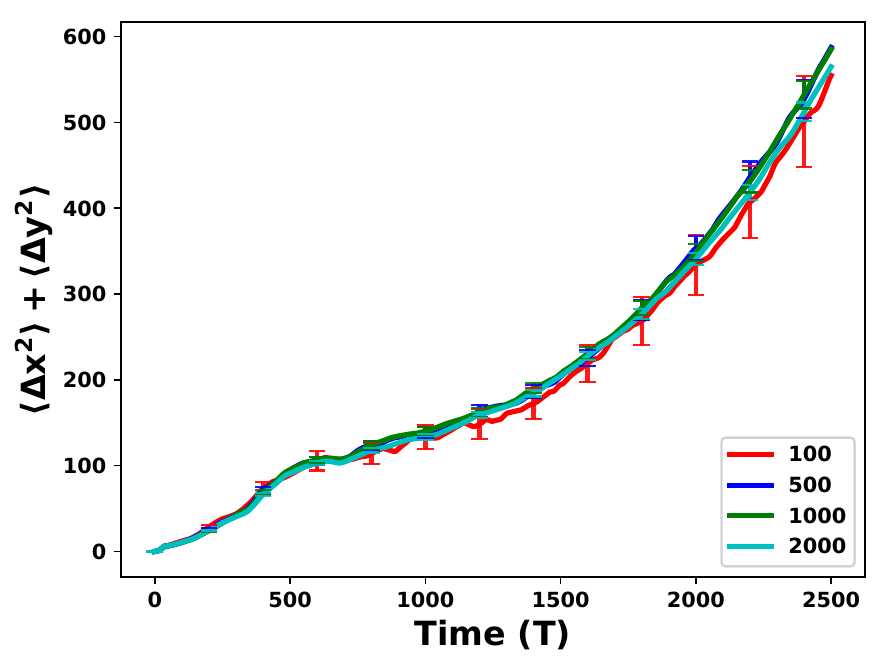}
        \caption{}
    \end{subfigure}
    \hfill
    \begin{subfigure}[h]{0.49\textwidth}
         \centering
         \includegraphics[width=\textwidth]{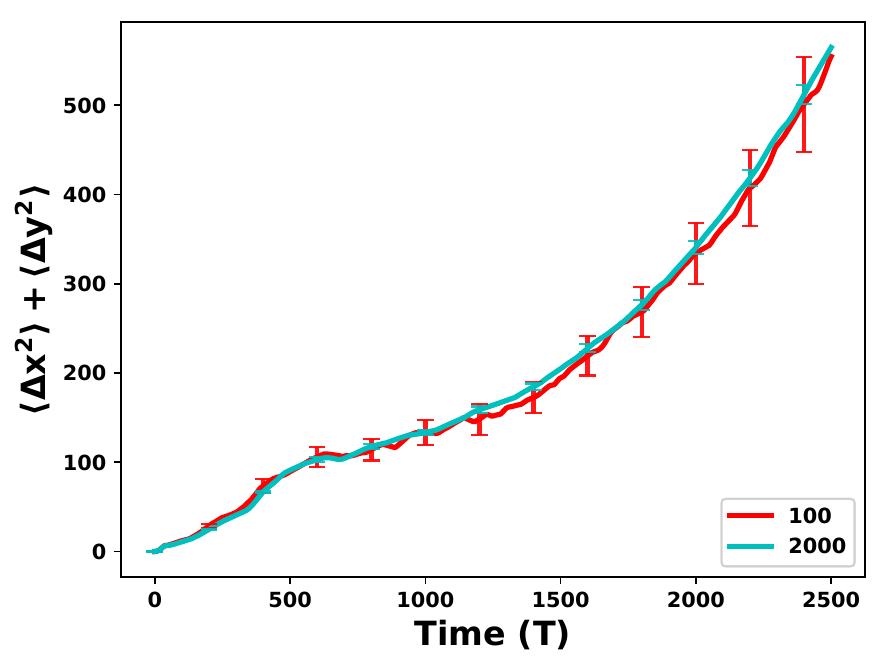}
        \caption{}
    \end{subfigure}
    \caption{{ {Statistical convergence of the mean-square displacement quantified using standard-error-of-the-mean (SEM) estimates. Results are shown for (a) all particle-number cases and (b) the extreme cases ($N_p$=100) and ($N_p$=2000). The differences between the MSD curves remain within the corresponding SEM bounds throughout the simulation, confirming statistical convergence for ($N_p \gtrsim $1000).}}}
        \label{fig:particle_convergence2}

\end{figure*}

To further quantify the statistical uncertainty associated with finite particle sampling, standard-error-of-the-mean (SEM) estimates are computed for the MSD. The resulting error bounds are shown in Fig.~\ref{fig:particle_convergence2}. The differences between the transport statistics obtained for different particle numbers remain within the corresponding SEM bounds throughout the simulation. These results demonstrate that the computed transport quantities are statistically converged for particle numbers ($N_p \gtrsim$ 1000).

Based on these convergence tests, all subsequent tracer-particle simulations reported in this work employ $N_p=1000$, which provides statistically converged transport measures while maintaining computational efficiency.

\section{Applications}\label{}

\subsection{Tracer particle dynamics in decaying two dimensional incompressible Navier-Stokes turbulence}

To demonstrate the capability of the coupled Eulerian--Lagrangian framework in dynamically evolving turbulent flows, tracer particles are evolved within a decaying two-dimensional incompressible Navier--Stokes simulation performed at a resolution of $2048^2$ and Reynolds number $\mathrm{Re}=228576$. The flow is initialized using alternating-sign vorticity strips and evolved from the Kelvin–Helmholtz instability through turbulence development and late-time coherent-vortex formation. Tracer particles are integrated simultaneously with the fluid evolution using the interpolation procedure described earlier. The simulation parameters correspond to those employed in our previous study of long-time turbulent transport [\cite{Maiti2026PoF}].

\begin{figure*}
\centering

\begin{subfigure}[b]{0.32\textwidth}
\centering
\includegraphics[width=\textwidth]{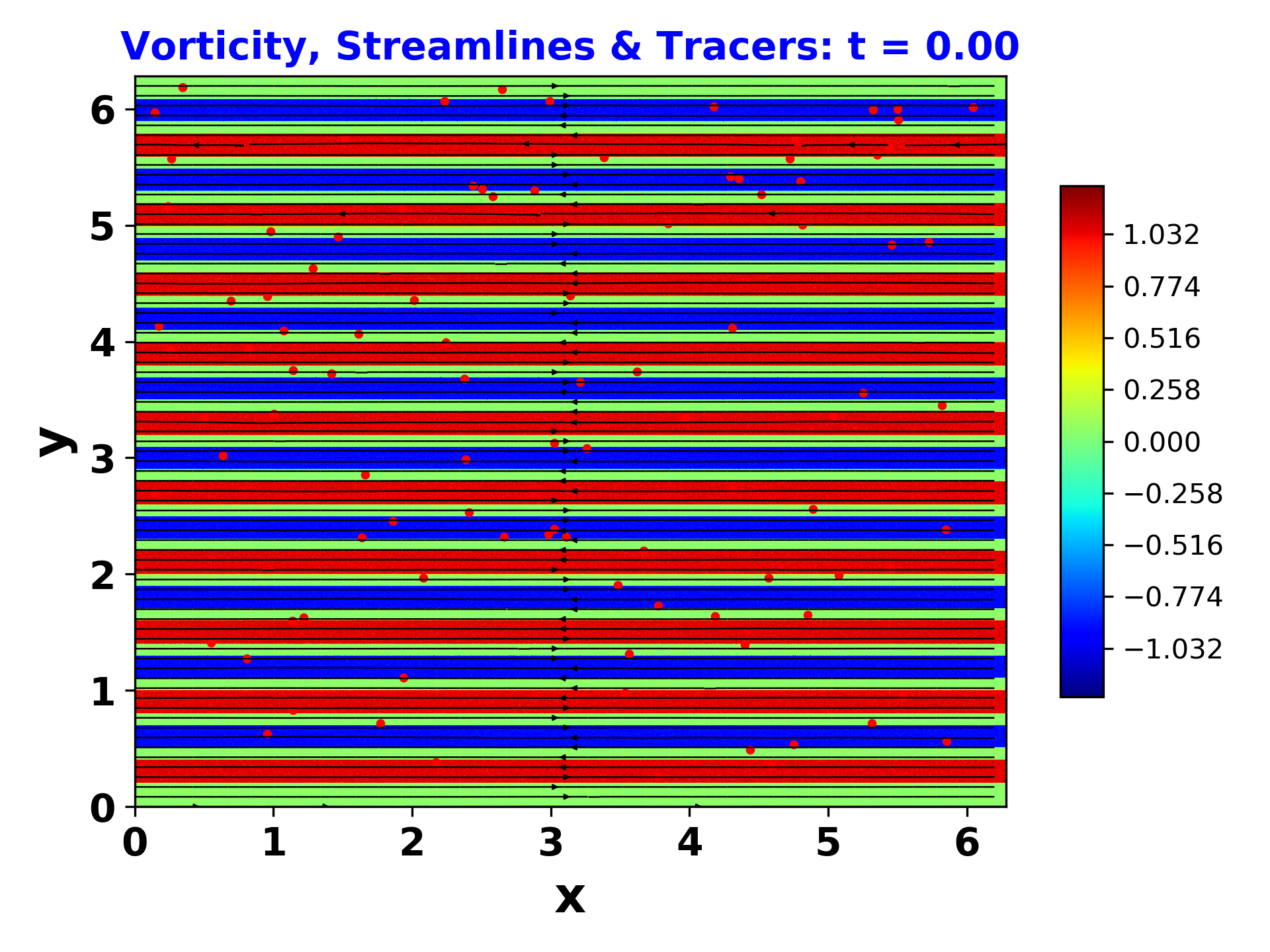}
\caption{$T=0$}
\end{subfigure}
\hfill
\begin{subfigure}[b]{0.32\textwidth}
\centering
\includegraphics[width=\textwidth]{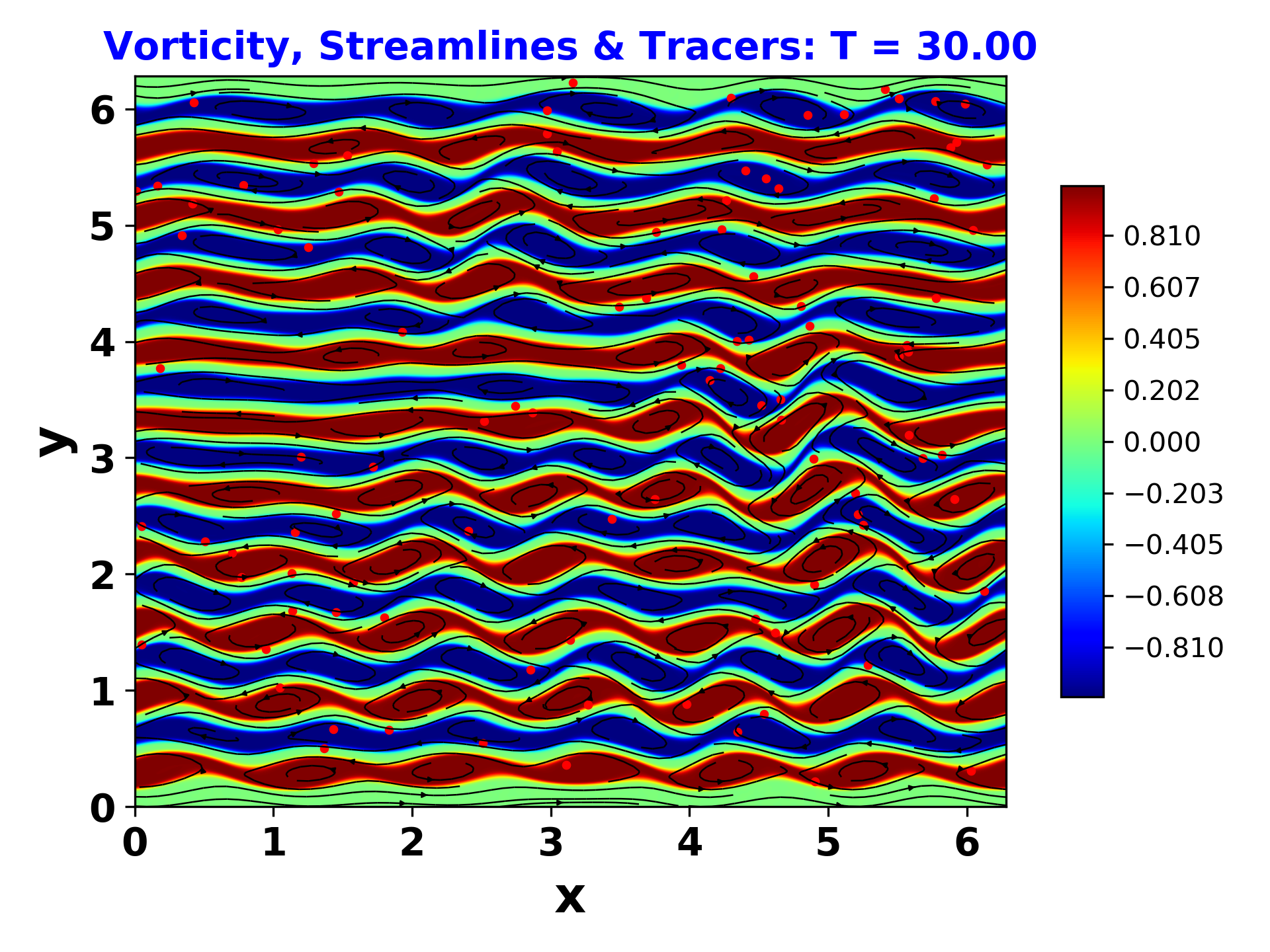}
\caption{$T=50$}
\end{subfigure}
\hfill
\begin{subfigure}[b]{0.32\textwidth}
\centering
\includegraphics[width=\textwidth]{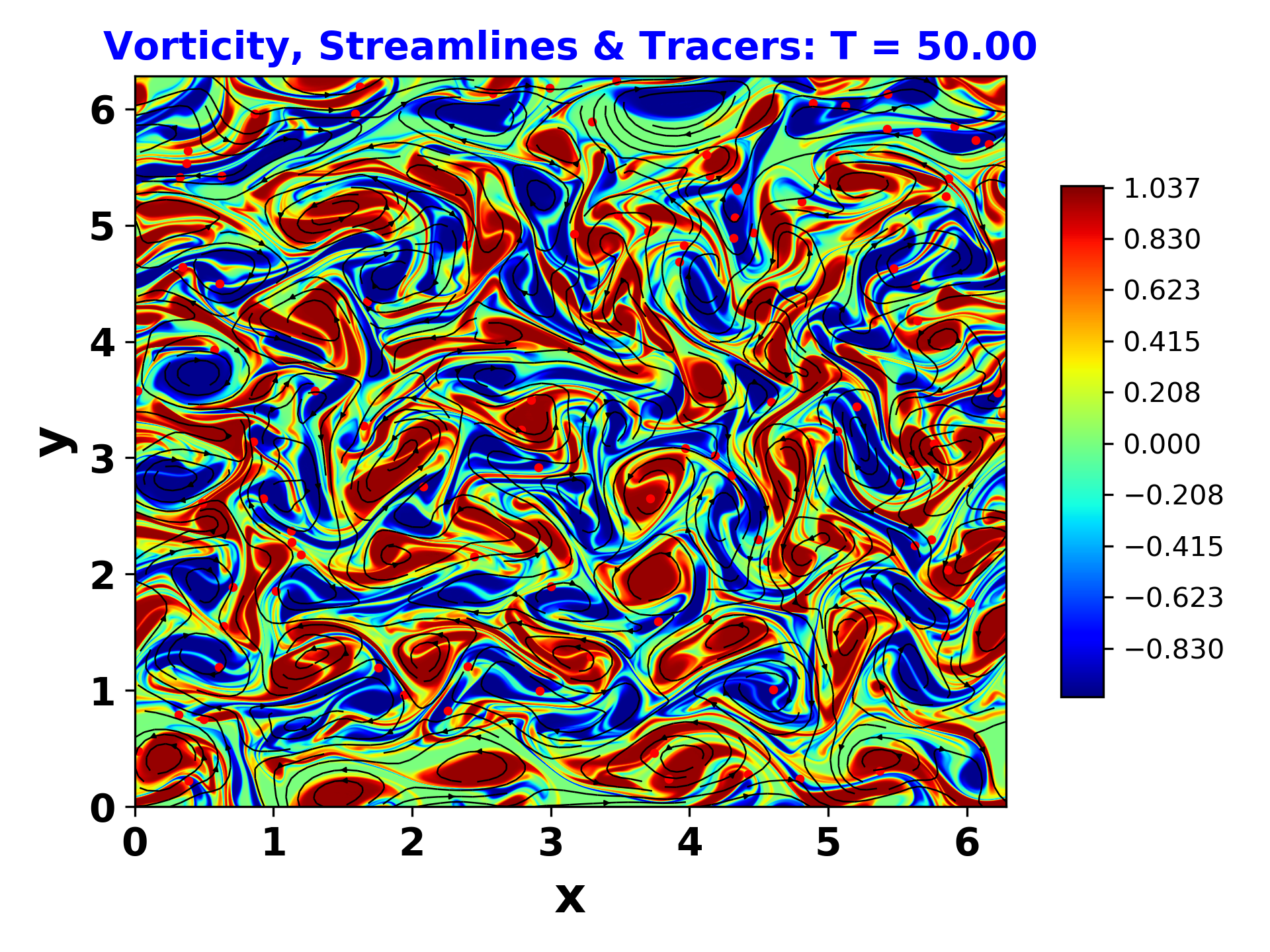}
\caption{$T=200$}
\end{subfigure}

\vspace{0.3cm}

\begin{subfigure}[b]{0.32\textwidth}
\centering
\includegraphics[width=\textwidth]{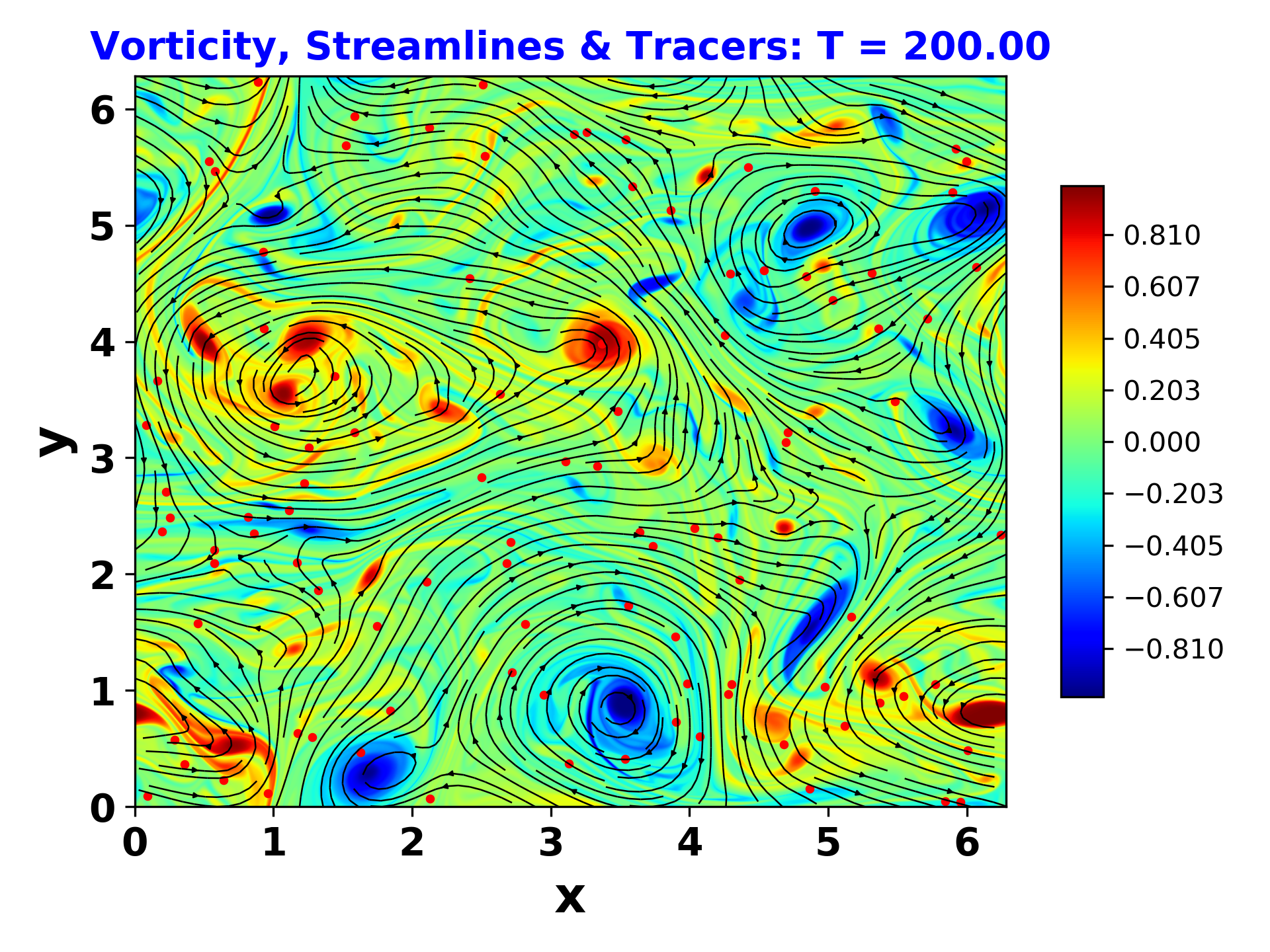}
\caption{$T=500$}
\end{subfigure}
\hfill
\begin{subfigure}[b]{0.32\textwidth}
\centering
\includegraphics[width=\textwidth]{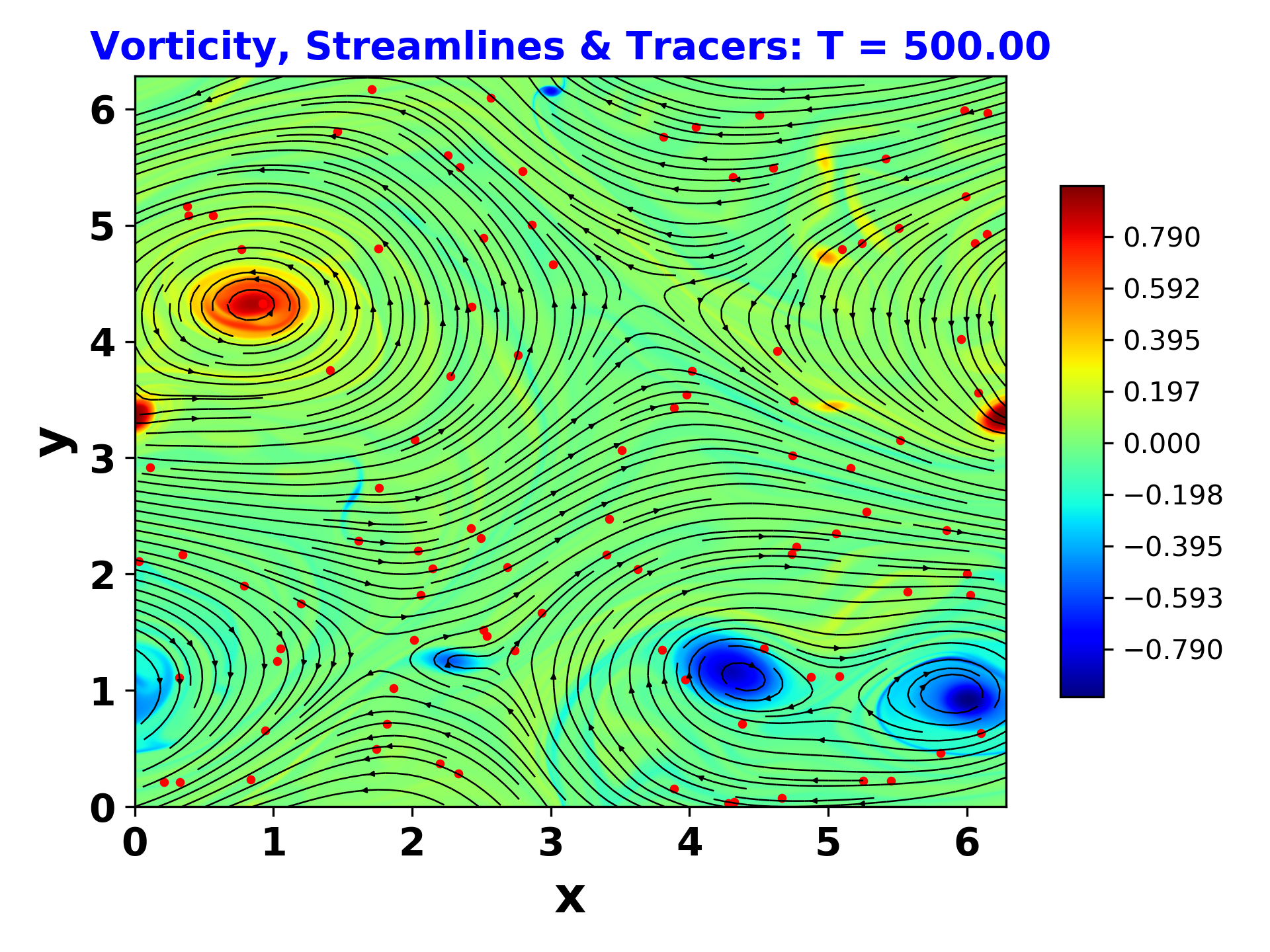}
\caption{$T=1000$}
\end{subfigure}
\hfill
\begin{subfigure}[b]{0.32\textwidth}
\centering
\includegraphics[width=\textwidth]{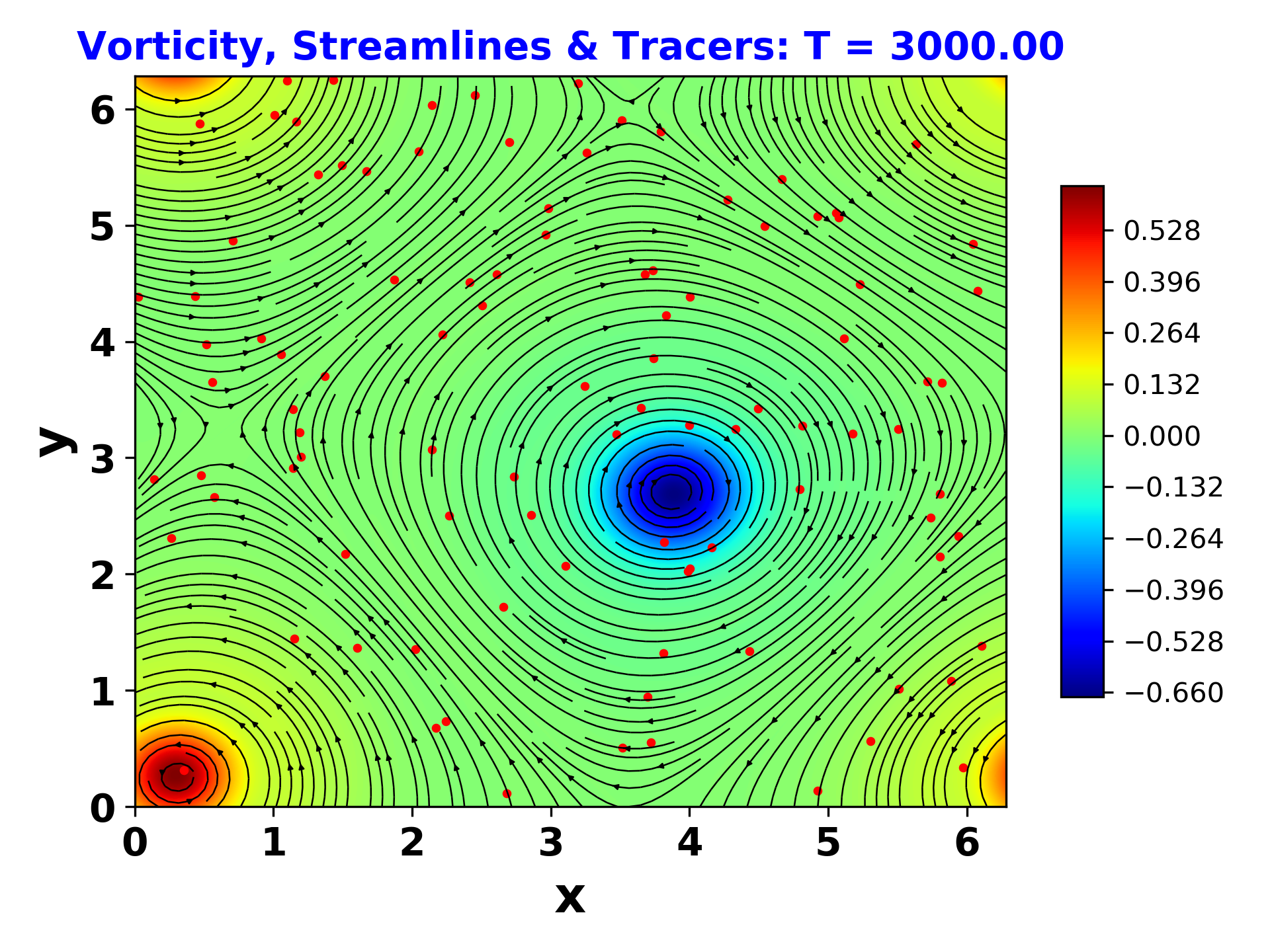}
\caption{$T=3000$}
\end{subfigure}

\caption{
Tracer transport in decaying two-dimensional incompressible turbulence for a 20-strip initial condition. The snapshots show the simultaneous evolution of the Eulerian flow field and the Lagrangian tracer particles from the initial Kelvin--Helmholtz instability through turbulence development and late-time coherent-vortex formation. The coupled Eulerian--Lagrangian framework accurately captures particle advection, vortex trapping, and long-time transport in dynamically evolving turbulent flows. Figure adapted from Ref.~[\cite{Maiti2026PoF}].
}
\label{fig:turbulent_transport}
\end{figure*}

Figure~\ref{fig:turbulent_transport} illustrates the evolution of tracer particles in a dynamically evolving turbulent flow. Following the onset of the Kelvin--Helmholtz instability, the initially ordered vorticity strips roll up into vortices, generating a complex turbulent velocity field that continuously deforms the tracer distribution. As the flow evolves, particles become entrained within coherent vortices, are transported through vortex interactions and mergers, and eventually follow the large-scale coherent structures that emerge during the late stages of the inverse cascade. The results demonstrate the ability of the coupled Eulerian--Lagrangian framework to accurately track large particle ensembles over long integration times while preserving the evolving flow topology.

\begin{figure*}
\centering
  \includegraphics[width=\textwidth]{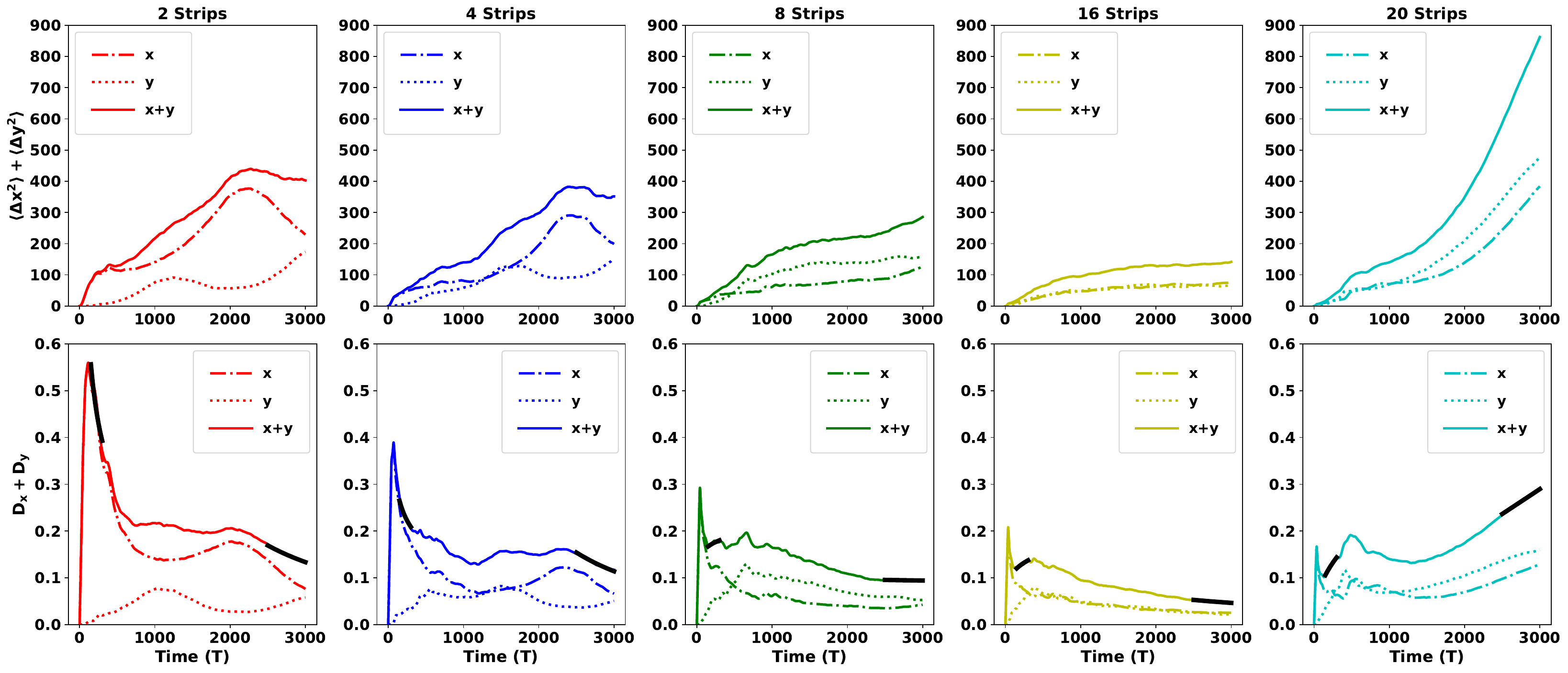}
 
\caption{
Tracer transport statistics obtained using the coupled Eulerian--Lagrangian framework for decaying two-dimensional turbulence initialized with different vorticity packing fractions (2, 4, 8, 16, and 20 strips). The upper panels show the evolution of the mean-square displacement, while the lower panels show the corresponding diffusion coefficients. The framework successfully resolves the dependence of particle transport on the underlying turbulent dynamics, including anisotropic-to-isotropic transport transitions, distinct diffusive regimes, and long-time particle dispersion across a broad range of flow configurations. Figure adapted from Ref.~[\cite{Maiti2026PoF}].
}

  \label{fig:tr_all_xy}
\end{figure*}

To further demonstrate the capability of the framework for quantitative transport studies, we compute the mean-square displacement (MSD) and diffusion coefficients for simulations initialized with different vorticity packing fractions or number of strips (2, 4, 8, 16, 20). The resulting transport statistics are shown in Fig.~\ref{fig:tr_all_xy}. The framework successfully captures the dependence of particle transport on the underlying turbulent dynamics, reproducing the transition between different transport regimes and resolving both anisotropic and isotropic transport behavior over extended simulation times. These results illustrate the suitability of the developed framework for investigating turbulent dispersion, anomalous transport, and long-time Lagrangian dynamics in two-dimensional incompressible flows.

Overall, these results demonstrate that the coupled Eulerian--Lagrangian framework can accurately track large ensembles of tracer particles over long integration times while simultaneously resolving the evolving turbulent flow field. The framework therefore provides a robust tool for investigating turbulent dispersion, anomalous transport, coherent-structure interactions, and long-time Lagrangian dynamics in incompressible flows.

\subsection{Inertial particle dynamics in decaying two dimensional incompressible Navier-Stokes turbulence}


To demonstrate the capability of the framework for finite-inertia particle dynamics, inertial particles are evolved simultaneously with decaying two-dimensional incompressible turbulence. Simulations are performed for a range of Stokes numbers, spanning the tracer limit to strongly inertial particles, and for different initial vorticity packing fractions (VPFs). The particle dynamics are obtained using the one-way coupled Eulerian--Lagrangian formulation described earlier, with fluid velocities interpolated to the instantaneous particle positions.

\begin{figure*}
\centering

\begin{subfigure}[b]{0.47\textwidth}
\centering
\includegraphics[width=\textwidth]{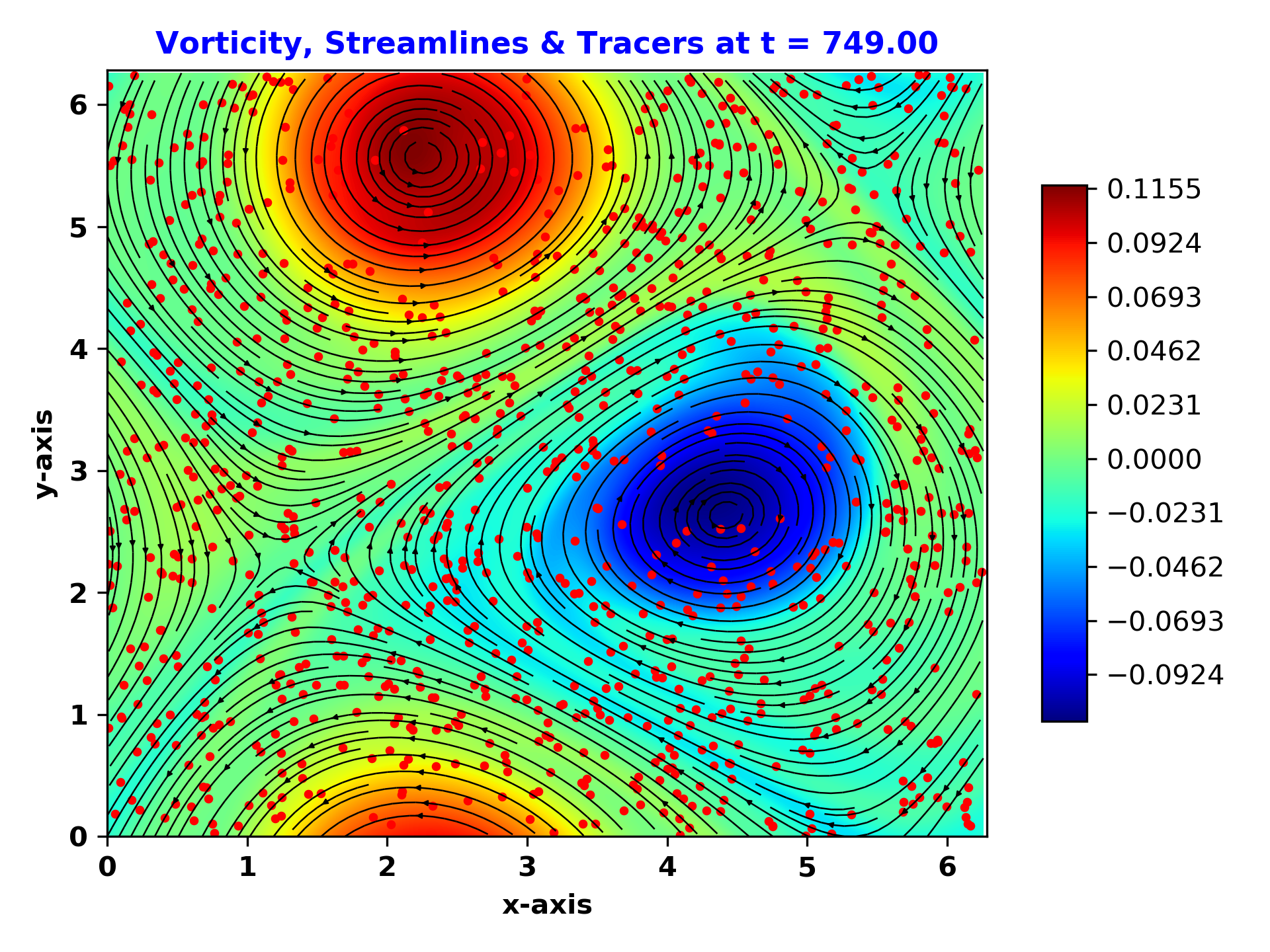}
\caption{St 0.2}
\end{subfigure}
\hfill
\begin{subfigure}[b]{0.47\textwidth}
\centering
\includegraphics[width=\textwidth]{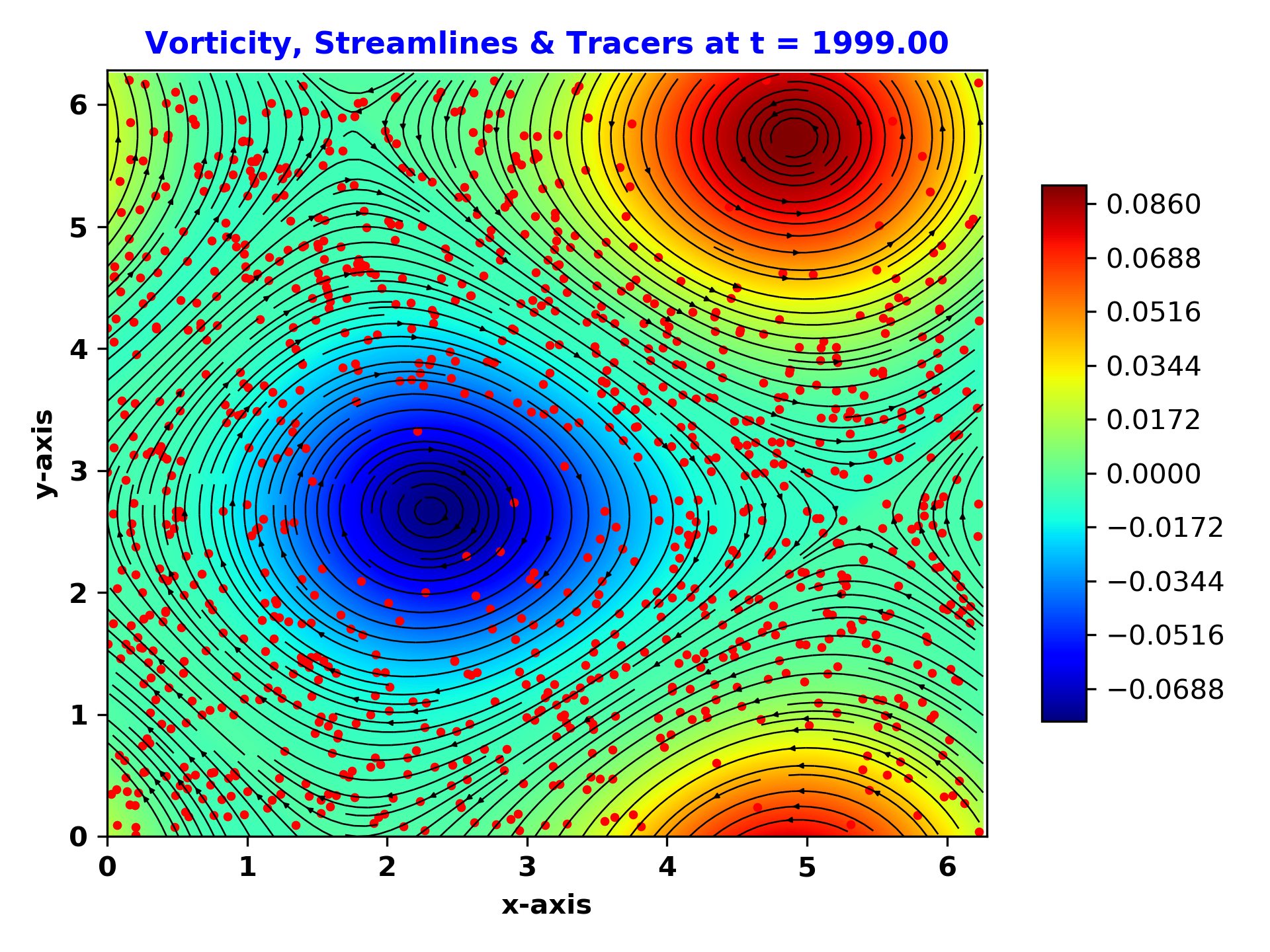}
\caption{St 0.2}
\end{subfigure}

\vspace{0.25cm}


\vspace{0.25cm}

\begin{subfigure}[b]{0.47\textwidth}
\centering
\includegraphics[width=\textwidth]{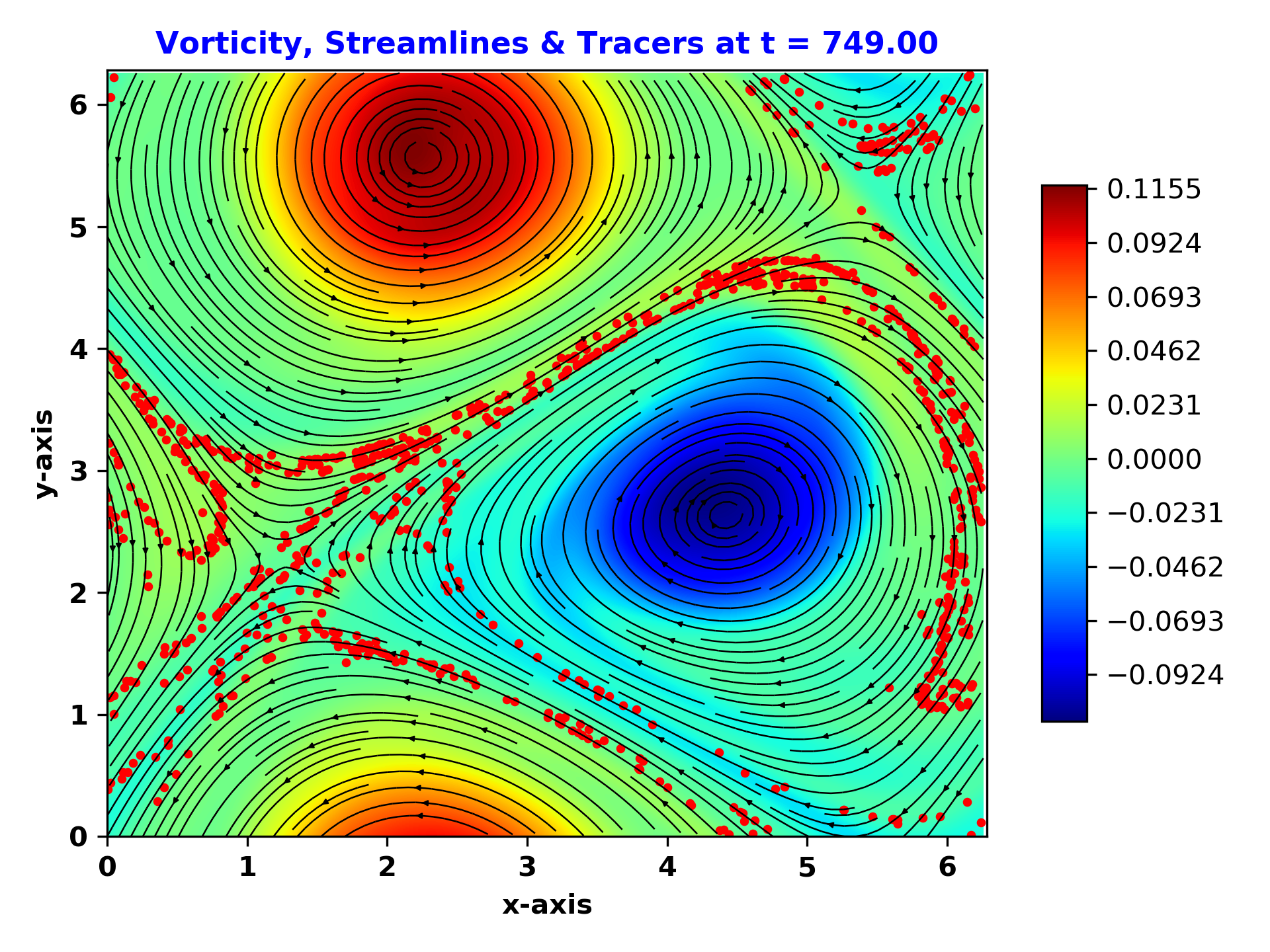}
\caption{St 10}
\end{subfigure}
\hfill
\begin{subfigure}[b]{0.47\textwidth}
\centering
\includegraphics[width=\textwidth]{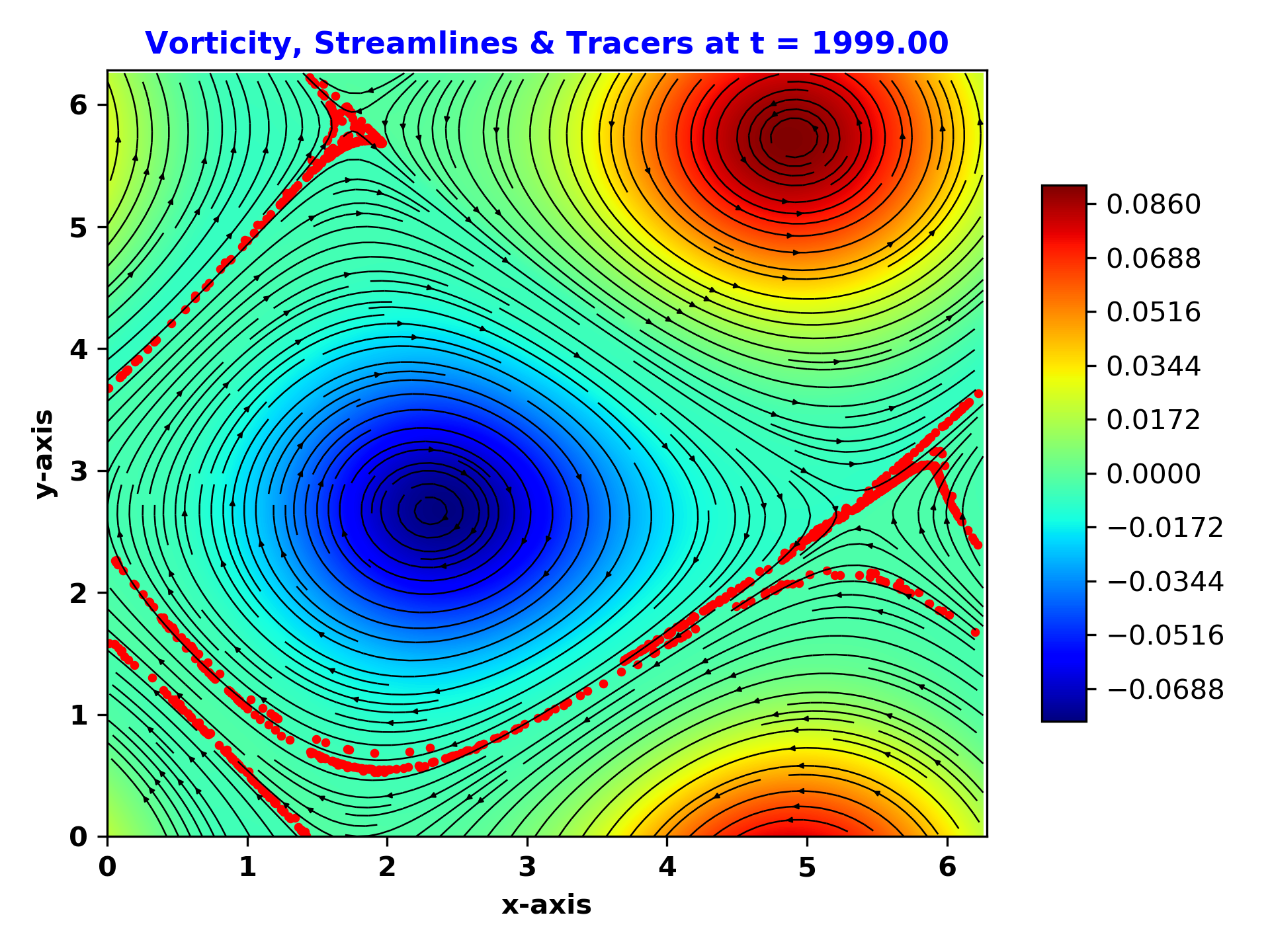}
\caption{St 10}
\end{subfigure}

\caption{
 Distribution of inertial particles in decaying two-dimensional incompressible turbulence for the 20-strip initial condition at two representative times, T = 749 and T = 1999. Results are shown for Stokes numbers St = 0.2, and 10. Weakly inertial particles (St = 0.2) remain broadly distributed throughout the flow and closely follow the underlying streamlines. As the Stokes number increases, preferential concentration becomes increasingly pronounced. Particles with higher St exhibit partial depletion from vortex cores and enhanced accumulation in strain-dominated regions, while strongly inertial particles (St = 10) form highly concentrated filamentary clusters and are largely excluded from coherent vortical structures. The snapshots demonstrate the ability of the Eulerian--Lagrangian framework to capture inertia-induced clustering and preferential concentration in dynamically evolving turbulent flows.
}
\label{fig:inertial_particles}
\end{figure*}

Figure~\ref{fig:inertial_particles} illustrates the spatial distribution of inertial particles for two representative Stokes numbers (St = 0.2 and 10) at two different stages of the turbulent evolution. A clear dependence of particle clustering on inertia is observed. For weakly inertial particles (St = 0.2), the particle distribution remains relatively homogeneous and closely follows the underlying flow streamlines, indicating strong coupling to the carrier fluid. As the particle inertia increases, preferential concentration becomes apparent, with particles becoming depleted from coherent vortex cores and accumulating in the surrounding strain-dominated regions. For strongly inertial particles (St = 10), clustering is significantly enhanced, producing narrow filamentary structures and highly non-uniform particle distributions. In this regime, the particle response time exceeds the characteristic turnover time of many turbulent structures, causing particles to decouple from the local fluid motion and concentrate along persistent transport pathways. These results demonstrate that the developed framework accurately reproduces the well-known inertia-induced preferential concentration phenomenon and can efficiently capture particle clustering over long turbulent integration times.

\begin{figure*}
\centering
  \includegraphics[width=\textwidth]{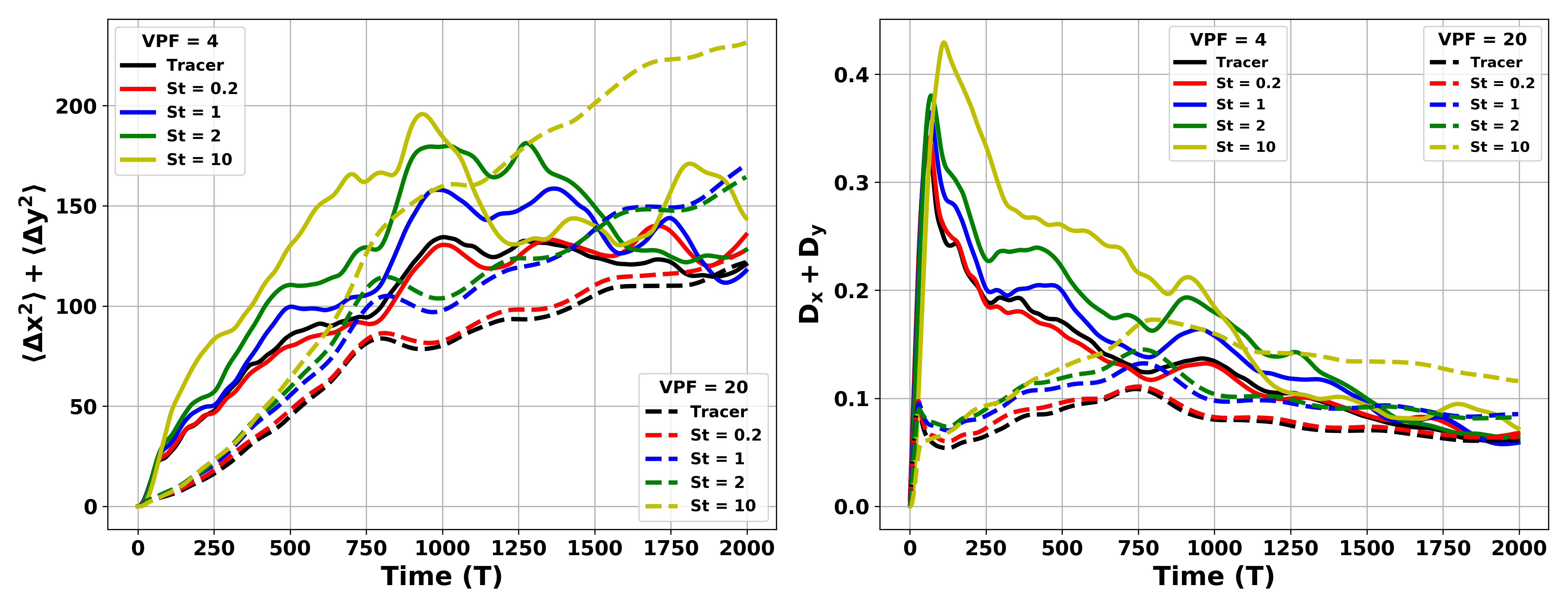}
 
\caption{Transport statistics of inertial particles in decaying two-dimensional incompressible turbulence for two representative vorticity packing fractions, VPF = 4 and VPF = 20. The left panel shows the evolution of the mean-square displacement (MSD), while the right panel shows the corresponding diffusion coefficient $(D_x+D_y)$. Results are presented for passive tracers and inertial particles with Stokes numbers ({St}=0.2, 1, 2, and 10. Increasing particle inertia generally leads to enhanced long-time transport, whereas the higher-vorticity-packing-fraction configuration exhibits systematically reduced dispersion compared to the lower-vorticity-packing-fraction case. The figure demonstrates the ability of the coupled Eulerian--Lagrangian framework to capture the combined effects of turbulence and particle inertia on transport.}
  
  \label{fig:inertial_transport}
\end{figure*}

Figure~\ref{fig:inertial_transport} shows the evolution of the mean-square displacement and the corresponding diffusion coefficients for representative VPF = 4 and VPF = 20 configurations. The framework successfully resolves the influence of particle inertia on long-time transport. In both flow configurations, particle dispersion generally increases with increasing Stokes number, with highly inertial particles exhibiting the largest mean-square displacements. The simulations also capture the dependence of transport on the underlying turbulent flow structure, with the lower-vorticity-packing-fraction case producing systematically larger particle transport than the higher-packing-fraction case.

The increase in transport with Stokes number can be attributed to the finite response time of inertial particles, which allows them to deviate from fluid streamlines and reduces trapping within coherent vortical structures. The reduced transport observed for larger vorticity packing fractions is associated with the more isotropic and densely interacting vortex population that develops in these flows, leading to enhanced particle mixing and reduced net displacement.

These results demonstrate that the developed framework can efficiently simulate large ensembles of finite-inertia particles in dynamically evolving turbulent flows and accurately capture the combined effects of fluid turbulence and particle inertia on transport processes.

\section{Performance Analysis}\label{}

\subsection{Resolution scaling of the coupled solver}

To quantify the computational performance of the coupled Eulerian--Lagrangian framework, simulations were performed on a single GPU while varying the spatial resolution of the Eulerian fluid solver. The number of particles ($N_p = 1000$), timestep size ($\Delta t = 10^{-3}$), and simulation duration ($T = 10$) were kept fixed in all cases in order to isolate the effect of grid resolution on the computational cost.

\begin{figure*}
\centering

\begin{subfigure}[b]{0.49\textwidth}
\centering
\includegraphics[width=\textwidth]{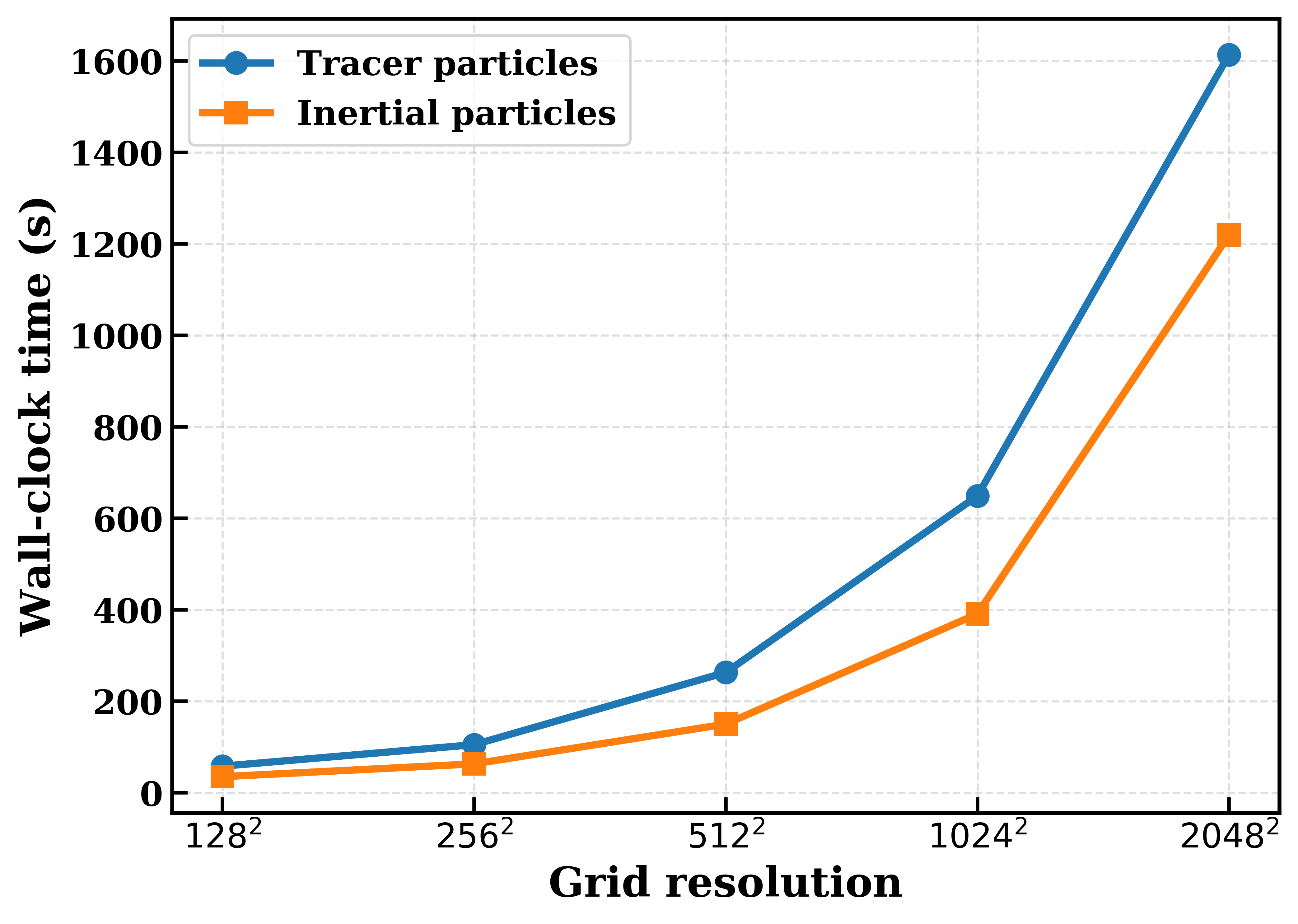}
\caption{Resolution scaling}
\label{fig:resolution_scaling}
\end{subfigure}
\hfill
\begin{subfigure}[b]{0.49\textwidth}
\centering
\includegraphics[width=\textwidth]{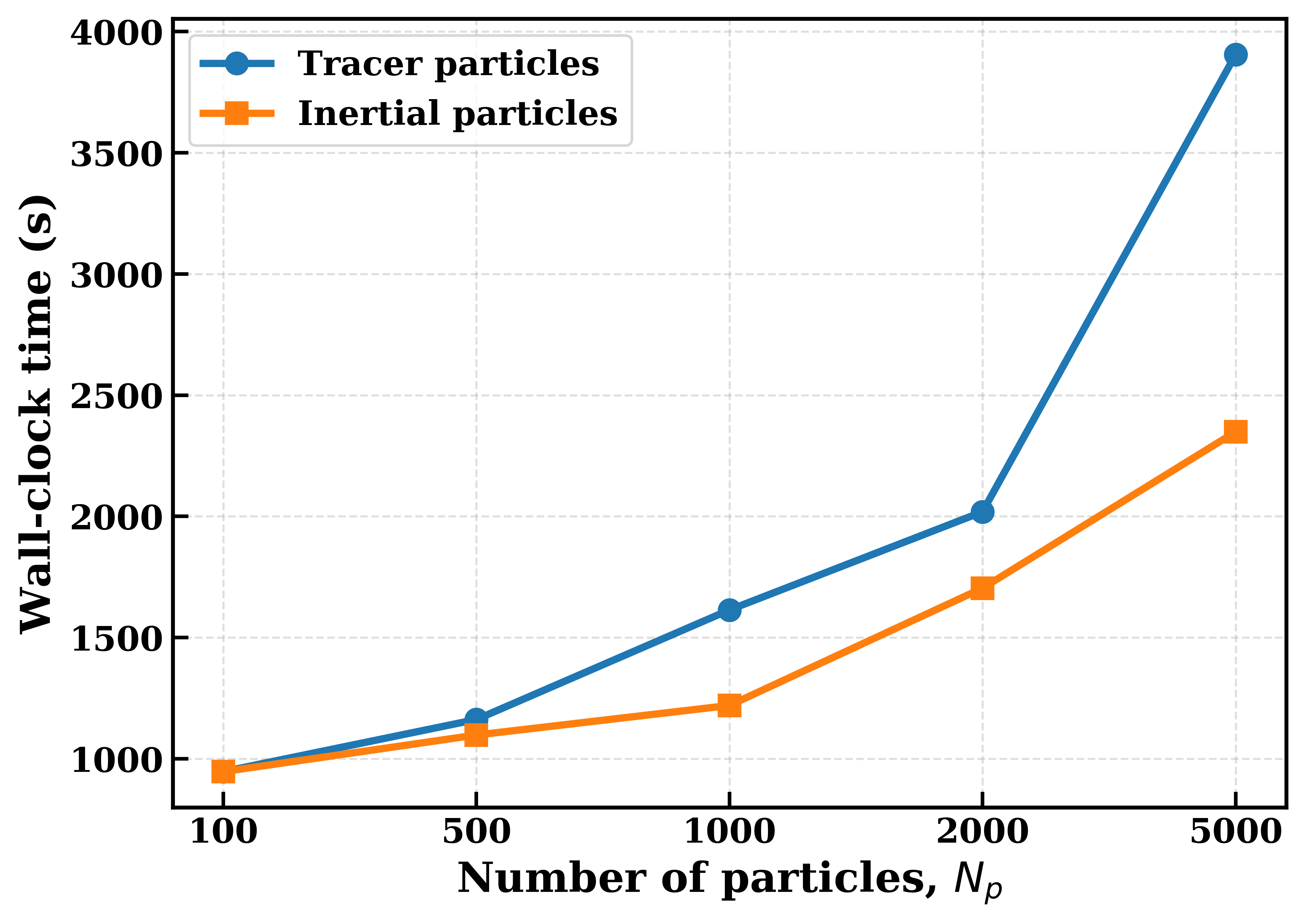}
\caption{Particle-number scaling}
\label{fig:particle_scaling}
\end{subfigure}

\caption{
Performance scaling of the coupled Eulerian--Lagrangian solver. 
(a) Wall-clock time as a function of Eulerian grid resolution for fixed particle number.
(b) Wall-clock time as a function of particle number at fixed grid resolution $2048^2$.
Results are shown for both tracer and inertial particle implementations.
}
\label{fig:performance_scaling}
\end{figure*}

Table~\ref{tab:resolution_scaling} summarizes the measured wall-clock times for grid resolutions ranging from $128^2$ to $2048^2$. The corresponding scaling behavior is shown in Fig.~\ref{fig:resolution_scaling}. As expected, the computational cost increases monotonically with increasing spatial resolution due to the larger number of grid points and the increased cost of the underlying pseudospectral operations.

\begin{table*}
\centering
\caption{Resolution scaling of the coupled Eulerian--Lagrangian solver.}
\label{tab:resolution_scaling}
\begin{tabular}{ccc}
\hline
\multirow{2}{*}{Grid resolution} &
\multicolumn{2}{c}{Wall-clock time (s)} \\
\cline{2-3}
 & Tracer particles & Inertial particles \\
\hline
$128^2$  & 58   & 35   \\
$256^2$  & 105  & 63   \\
$512^2$  & 263  & 150  \\
$1024^2$ & 649  & 391  \\
$2048^2$ & 1614 & 1220 \\
\hline
\end{tabular}
\end{table*}

For both tracer- and inertial-particle simulations, the runtime exhibits systematic growth with increasing grid resolution, demonstrating stable and predictable scaling of the coupled Eulerian--Lagrangian implementation over a broad range of problem sizes. The results further indicate that the framework remains computationally tractable at production resolutions of $2048^2$, which are routinely employed in the turbulence simulations presented in Sec.~6.

The measured wall-clock times include both the GPU-accelerated Eulerian fluid solver and the Lagrangian particle-tracking module, thereby reflecting the total computational cost of the coupled framework. The inertial-particle implementation is consistently less computationally expensive than the tracer-particle implementation in the present framework. This behavior arises from the different interpolation strategies employed by the two modules: the tracer solver interpolates the stream-function field and subsequently reconstructs the velocity components, whereas the inertial solver directly interpolates the fluid velocity field at the particle location, thereby reducing the interpolation overhead.

\subsection{Particle-number scaling of the coupled solver}

To quantify the computational overhead associated with particle tracking, simulations were performed at a fixed grid resolution of $2048^2$ while varying the number of particles. The timestep size ($\Delta t = 10^{-3}$), simulation duration ($T = 10$), and all other numerical parameters were kept unchanged in order to isolate the effect of particle number on the computational cost. The measured wall-clock times for both tracer- and inertial-particle simulations are summarized in Table~\ref{tab:particle_scaling}.

\begin{table*}
\centering
\caption{Particle-number scaling of the coupled Eulerian--Lagrangian solver at fixed grid resolution ($2048^2$).}
\label{tab:particle_scaling}
\begin{tabular}{ccc}
\hline
\multirow{2}{*}{Number of particles, $N_p$} &
\multicolumn{2}{c}{Wall-clock time (s)} \\
\cline{2-3}
& Tracer particles & Inertial particles \\
\hline
100  & 947  & 947  \\
500  & 1162 & 1098 \\
1000 & 1614 & 1220 \\
2000 & 2019 & 1704 \\
5000 & 3905 & 2349 \\
\hline
\end{tabular}
\end{table*}

As expected, the total runtime increases with increasing particle number due to the additional interpolation and particle-integration operations performed at every timestep. However, the increase in runtime is significantly weaker than the corresponding increase in particle population. Increasing the number of particles from $N_p=100$ to $N_p=5000$, corresponding to a fifty-fold increase in particle count, increases the wall-clock time from 947 s to 3905 s for tracer particles and from 947 s to 2349 s for inertial particles.

These results indicate that the computational cost of the coupled Eulerian--Lagrangian framework remains largely dominated by the Eulerian fluid solver and its associated pseudospectral operations, while the Lagrangian particle module introduces a comparatively moderate overhead. Consequently, large particle ensembles containing several thousand particles can be incorporated into simulations without a prohibitive increase in computational cost.

A comparison of the tracer and inertial results further shows that the inertial-particle module introduces a smaller computational overhead than the tracer-particle implementation. Despite the additional evolution of particle velocities required by the inertial equations, the direct interpolation of the velocity field results in a lower overall computational cost than the stream-function-based interpolation and velocity reconstruction employed by the tracer solver.

Overall, these results demonstrate that the present implementation is well suited for long-time turbulent transport studies involving large particle populations while maintaining computational efficiency at production-scale resolutions.

\section{Conclusions}\label{}

In this work, we have developed, implemented, and verified a modular one-way coupled Eulerian--Lagrangian framework for simulating tracer and inertial particle dynamics in two-dimensional incompressible Navier--Stokes turbulence. The framework extends the GPU-accelerated GHD2D pseudospectral solver by incorporating Lagrangian particle modules within a unified computational environment, enabling the simultaneous evolution of the Eulerian flow field and large ensembles of passive tracer and finite-inertia particles. The implementation combines a Fourier pseudospectral fluid solver with fourth-order Runge--Kutta particle integration, synchronized through a one-way Eulerian--Lagrangian coupling strategy employing spatial and temporal interpolation of the carrier-flow quantities. In addition, the modular software architecture and heterogeneous GPU--CPU implementation provide a flexible foundation for future algorithmic extensions while retaining compatibility with the underlying Eulerian solver.

The numerical correctness and robustness of the framework have been established through a comprehensive hierarchy of verification tests. The Eulerian fluid solver was validated against the analytical growth rates of the Kelvin--Helmholtz instability, while the tracer-particle module reproduced the benchmark transport characteristics of a chaotic kinematic flow. The inertial-particle implementation was verified using both analytical solutions in a uniform flow and previously reported benchmark dynamics in steady Taylor--Green vortices. The interpolation infrastructure was systematically assessed through comparisons of bilinear, bicubic Catmull--Rom, and bicubic B-spline interpolation schemes, demonstrating that all three methods produce nearly identical transport statistics for the turbulent flows considered in this work. Consequently, the computationally efficient bilinear interpolation was adopted for all production simulations. Furthermore, particle-number convergence studies confirmed that statistically converged transport measures are obtained using approximately one thousand particles, thereby establishing the numerical reliability of the framework for long-time Lagrangian transport calculations.

The capabilities of the coupled solver were demonstrated through applications to dynamically evolving decaying two-dimensional turbulence. The framework accurately captures tracer-particle advection, vortex trapping, coherent-structure interactions, long-time turbulent dispersion, and the dependence of transport on the underlying turbulent flow evolution. For finite-inertia particles, the simulations successfully reproduce inertia-induced preferential concentration, particle clustering, and the influence of the Stokes number on long-time transport. These applications demonstrate that the framework provides a robust and versatile computational tool for investigating a broad range of Eulerian--Lagrangian transport phenomena in turbulent flows while maintaining accurate synchronization between the evolving carrier flow and the particle dynamics.

Performance measurements further demonstrate that the computational cost of the coupled framework exhibits stable and predictable scaling with both Eulerian grid resolution and particle number. The GPU-accelerated pseudospectral fluid solver remains the dominant computational component, while the additional cost associated with Lagrangian particle tracking constitutes only a moderate overhead, allowing production simulations at resolutions up to $2048^2$ with thousands of particles to be performed efficiently. The modular organization of the code further enables the independent development of the fluid solver, interpolation procedures, particle modules, and diagnostic components without altering the overall framework.

The present implementation adopts a one-way coupling approximation in which the carrier flow influences the particle motion while particle feedback on the fluid is neglected. Owing to its modular design, the framework provides a flexible foundation for future developments, including two-way particle--fluid coupling, additional particle-force models, alternative interpolation strategies, GPU acceleration of the Lagrangian particle modules, and extensions to three-dimensional, compressible, or magnetohydrodynamic flows. The computational framework developed in this work therefore provides a robust, efficient, and extensible platform for future studies of Eulerian--Lagrangian transport in turbulent fluid and plasma systems.

\section*{Acknowledgements}

The simulations and visualizations presented here are
performed on GPU nodes and visualization nodes of the ANTYA cluster at the Institute for Plasma Research (IPR), India. The authors are grateful to the HPC support team of IPR for extending their help related to the ANTYA cluster. The authors are also grateful to Prof. Guglielmo Lacorata of the Institute of Marine Sciences (CNR–ISMAR), Rome, for his valuable insights in benchmarking the particle solver used in this study against  results from the kinematic 2D chaotic flow investigated by his group. The authors gratefully acknowledge Dr.~Rupak Mukherjee for developing the original GHD2D fluid solver and Dr.~Shishir Biswas for developing its GPU-accelerated implementation and for kindly providing the source code upon which the present Eulerian--Lagrangian framework has been developed.

\section*{Declaration of competing interest}

The author declares that there are no known competing financial interests or personal relationships that could have appeared to influence the work reported in this paper.

\section*{Data availability}

The data that support the findings of this study are available from the corresponding author upon reasonable request.

\printcredits
Conceptualization, Methodology, Software, Validation, Formal analysis, Investigation, Visualization, Writing -- original draft, Writing -- review \& editing.

\textbf{Rajaraman Ganesh:}  Supervision, Methodology, Investigation, Resources,  Writing -- review \& editing.
\bibliographystyle{cas-model2-names}

\bibliography{cas-refs}



\end{document}